\documentstyle[eqsecnum,aps,epsfig]{revtex}
\begin{document}
\draft
\title{Shell Model for Time-correlated Random Advection of Passive
  Scalars} 
\author{K. H. Andersen} 
\address{Niels Bohr Institute,
  Blegdamsvej 17, DK-2100 Copenhagen, Denmark {\em and} Institute of
  Hydraulic Research and Water Resources, The Danish Technical
  University, DK-2800 Lyngby, Denmark. e-mail: ken@isva.dtu.dk}
\author{P. Muratore-Ginanneschi} 
\address{Niels Bohr Institute,
  Blegdamsvej 17, DK-2100 Copenhagen {\O}, Denmark, e-mail: pmg@nbi.dk}
\date{\today}
\maketitle
\begin{abstract}
  We study a minimal shell model for the advection of a passive scalar
  by a Gaussian time correlated velocity field.  The anomalous scaling
  properties of the white noise limit are studied analytically.  The
  effect of the time correlations are investigated using perturbation
  theory around the white noise limit and non-perturbatively by
  numerical integration. The time correlation of the velocity field is
  seen to enhance the intermittency of the passive scalar.
\end{abstract}
\pacs{PACS numbers: 47.27.Gs, 47.27.Jv}

\section{Introduction}
\label{s:intro}
The advection of a scalar observable $\theta(x,t)$ by a velocity field 
$\bf{v}$ is described in classical hydrodynamics by the linear PDE 
\begin{equation}
\partial_t \theta + {\bf v}\cdot \nabla \theta =\kappa \nabla^2 \theta+{\bf f}.
\label{pde}
\end{equation} 
If $\bf{v}$ is assumed to be solution of the Navier-Stokes equations in a 
turbulent r\'egime and the P\'eclet number $Pe$, which measures the ratio 
between the strength of the advective effects and the molecular diffusion 
$\kappa$ in (\ref{pde}), is large
\begin{eqnarray}
Pe\equiv \frac{L v}{\kappa} \gg 1
\nonumber
\end{eqnarray}
($L$ and $v$ are the characteristic length and advection velocity in
the problem) and if a steady state is reached, an inertial range sets
in where both the effects of the forcing $f$ limited to the large
scales and those of the molecular diffusion acting mainly on the small
scales can be neglected.  In the inertial range no typical scale is
supposed to characterise the flow.  As a consequence, the structure
functions of the scalar field
\begin{equation}
S_{p}({\bf r})=\left<[\theta({\bf x}+{\bf r})-\theta({\bf x})]^p\right>
\label{structurefunc}
\end{equation} 
display a power law behaviour in the inertial range with anomalous
scaling exponents $H(p)$ \cite{Frisch}. The word anomalous means that
the exponents $H(p)$ deviates from the linear behaviour predicted by a
direct scaling analysis of (\ref{pde}).

It was first realised by Kraichnan \cite{Kraichnan} that anomalous
scaling can be observed in the mathematically more tractable case of
the advection by a random homogeneous and isotropic Gaussian velocity
field, which is delta correlated (white noise) in time and has zero
average and covariance in $d$ dimensions given by:
\begin{eqnarray}
\left<{\bf v}_i({\bf x},t){\bf v}_j({\bf y},s)\right>=\delta(t-s)
\left[D_{i,j}(0)-
D_{0}|{\bf x-y}|^{\xi_{wn}} \left(d-1+\xi_{wn}\right)\delta_{i,j}+\xi_{wn} 
\frac{(x-y)_{i}\,(x-y)_{j}}{|{\bf x-y}|^2}\right].
\nonumber
\end{eqnarray}
The power law behaviour of the covariance mimics an infinite inertial
range for the velocity field. The scaling exponent $\xi_{wn}$ is a
free parameter characterising the degree of turbulence of the
advecting field. The physically meaningful values range from zero to
two. In the first limit the effect of the random advection is just to
define an effective diffusion constant \cite{BeGaKu}.  In the latter
case the velocity increments are smooth as it is expected for a
laminar flow. The choice $\xi_{wn}$ equal to four thirds represents
the scaling of the velocity field conjectured by Kolmogorov for the
solution of the Navier-Stokes equation in turbulent r\'egime.

The hypothesis of delta correlation in time is of great mathematical
advantage for it allows to write the equation of motions of the scalar
correlations in a linear closed form.  The evolution of each
correlation in the inertial range is specified by a linear
differential operator, the inertial operator, plus matching conditions
at the boundary of the inertial range.  The occurrence of anomalous
scaling has been related to the existence of zero modes of the
inertial operators dominating the scaling properties of higher order
correlations (\cite{GaKu,CFKL,BeGaKu,CF} and for a recent review and
more complete bibliography \cite{Ga}).  The behaviour of the anomaly 
has also been numerically measured for the fourth order structure function 
versus the turbulence parameter $\xi_{wn}$ \cite{FMV}. 
However, to implement accurate numerical experiments still
remains a difficult task. Therefore it turns out to be useful to use
shell model as laboratories to test ideas and results related to the
full PDE model (see \cite{libro} for a general introduction to the
shell model concept). In \cite{WB,BBW} two different shell models
advected by a delta correlated velocity field mimicking the Kraichnan
model were constructed. Anomalous scaling was observed numerically and 
in the simpler case \cite{BBW} it was proven analytically that the
anomaly of the fourth order structure function is related to the
anomalous scaling of the dominant zero mode of the inertial operator.

The passive scalar advection by a white noise velocity field is a
useful mathematical model, but it still very far from a physical
realistic velocity field which displays both time correlations and
non Gaussian fluctuations. A first small step in this direction is
made by investigating how the introduction of a time correlation in a
Gaussian velocity field affects the statistical properties of the
scalar field.

In the present paper we introduce a time correlated velocity field in
a shell model. This is done by substituting the white noise with the
Ornstein-Uhlenbeck process which provides exponentially decaying time
correlations (Section~\ref{s:model}).  We investigate the model both
analytically and numerically.  By means of stochastic variational
calculus, which we shortly review in Appendices \ref{ap:one} and
\ref{ap:two}, we show how to rewrite the equations of motion for the
scalar correlations in integral non closed form. Such an operation
allows the evaluation of the correction to the white noise inertial
operator stemming from the time correlated velocity field. This
procedure has the further advantage that it creates a non-ambiguous
relation between the coupling terms for the scaling exponent
$\xi_{wn}$ of white noise advection to the scaling exponent $\xi$ of
the equal time correlation of the time-correlated velocity field
(Section~\ref{s:closure}).

The inertial operators can be expanded around the white noise limit in
powers of an a-dimensional parameter which is interpreted as
proportional to the ratio $\epsilon$ between the time correlation and
the turn-over time of the advecting field.  We focus on the features
of the steady state. There we assume that the averages over the
Ornstein-Uhlenbeck process of all the observables are time
translational invariant. As a consequence the inertial operators
become linear up to any finite order in $\epsilon$.

In the white noise case, when $\epsilon$ is equal zero, we generalise
the procedure first introduced in \cite{BBW} and we show that the
scaling of the zero modes of the inertial operator of any order is
captured by focusing on nearest-shell interactions. The equations are
closed with a scaling Ansatz (section \ref{s:wnclosure}) by
postulating that the scalar field is ``close'' to a multiplicative
process.  Furthermore we perturb the closure scheme in order to
extract the first order corrections in $\epsilon$ to the anomalous
exponents for different values of $\xi$ ranging from zero to two.  The
prediction of perturbation theory is an $\epsilon$ dependence
(non-universality) of the exponents except for the second order $H(2)$
(Section \ref{s:perturbative}). The overall result is analogous to the 
one obtained in \cite{Chertkov} where a Gaussian time correlated velocity 
field is considered for the advection of the scalar field in (\ref{pde}):
the introduction of time correlation is seen to enhance intermittency.
The anomalies vanish smoothly in the laminar limit $\xi$ equal two.

To examine the validity of the results from the analytical
calculations and explore the regime with long time-correlations
($\epsilon \gg 0$) we turned to numerical experiments. The occurrence of
corrections to the anomalies predicted by the perturbation theory for
small values of $\epsilon$ was confirmed. However, strong
non-perturbative effects sets in and are dominating when the expansion 
parameter becomes of the order of unity.

\section{The model}
\label{s:model}

The model is defined by the equations $(m=1,2,...,N)$
\begin{eqnarray}
\label{passive}
  &&[\frac{d}{dt}+\kappa k_{m}^{2}\,]\theta_{m}(t)-\delta_{1m}f(t)=i[k_{m+1}
  \theta_{m+1}^{*}(t)u_{m}^{*}(t)-k_{m}\theta_{m-1}^{*}(t)u_{m-1}^{*}(t)]\\
  \label{ousde}
  &&u_{m}(t)=\frac{v_{m}}{\epsilon \sqrt{\tau_{m}}}\int_{0}^{t}ds\,
  e^{-\frac{t-s}{\epsilon \tau_{m}}}\eta_{m}(s)\\
\label{force}
&&f(t)=\frac{\tilde{f}}{\epsilon\sqrt{\tau}} 
\int_{0}^{t}ds\,e^{-\frac{t-s}{\epsilon \tau}}\eta(s)
\end{eqnarray}
where the star denotes complex conjugation, the $\eta_{m}(t)$'s and
$\eta(t)$ are independent white noises with zero mean value and
correlation:
\begin{equation}
  \left<\eta_{m}(t)\eta_{n}^{*}(s)\right> = 2 \delta_{mn}\delta(t-s) \quad 
  \mbox{and} \quad
  \left<\eta(t)\eta^{*}(s)\right>=2 \delta(t-s)
\end{equation}
The boundary conditions are $\theta_{0}=\theta_{N+1}=0$. 
The model might be regarded as a severe truncation of the equation of the 
passive scalar (\ref{pde}) in Fourier space. The field component $\theta_m$
is the representative of all the Fourier modes in the shell with a wavenumber 
ranging between $k_m=k_{0}\,\lambda^m$ and $k_{m+1}=k_{0}\,\lambda^{m+1}$.
The parameter $\lambda$ is the ratio between two adjacent scales and it is 
usually taken equal to two in order to identify each shell with an octave of 
wave numbers.
The energy transfer in a turbulent flow is conjectured to occur mainly 
through the interactions of eddies of the same size. As a consequence the 
interactions in Fourier space are assumed to be local. The ``localness'' 
conjecture \cite{Frisch} is the motivation for the restriction to 
nearest neighbours of the couplings among the shells.

In the absence of external forcing and dissipation the total ``energy'' of 
the passive field is conserved:
\begin{equation}
  \frac{d}{dt}\,E=\frac{d}{dt}\sum_{m=1}^{N} |\theta_m|^{2}=0 \quad 
  \mbox{for} \quad f(t)=\kappa=0.
  \label{energy}
\end{equation}
Far from the infra-red and the ultra-violet boundaries (i.e. for $1
\ll m \ll N$) the conservation of energy is expected to hold
approximately giving rise to an inertial range.  Equations
(\ref{ousde}) and (\ref{force}) describe the random evolution
according to Ornstein-Uhlenbeck (O-U) processes respectively of the
advecting and external force field. The O-U process has differentiable
realisations thus making the random differential equations with
multiplicative noise which specify the dynamics of the scalar $\theta$
independent of the discretisation prescription.

The velocity correlations are for $t \ge s$ 
\begin{equation}
  \left<u_{m}(t)u_{m}^{*}(s)\right>=
  \frac{|v_{m}|^{2}}{\epsilon}(e^{-\frac{t-s}{\epsilon 
  \tau_{m}}}-e^{-\frac{t+s}{\epsilon \tau_{m}}}).
\label{correlation}
\end{equation}
In the limit of large $t$ only the stationary part survive.
The a-dimensional parameter $\epsilon$ appearing in the definition of the O-U
processes (\ref{ousde}) and (\ref{force}) defines the strength of the time 
correlation in units of the typical times $\tau_m$. 
In the white noise limit one has
\begin{equation}
  \lim_{\epsilon \downarrow 0} \left<u_{m}(t)u_{m}^{*}(s)\right> =2
  |v_{m}|^{2}\delta(\frac{t-s}{\tau_{m}}).
  \label{whitenoise}
\end{equation}
The factor two is reintroduced here with the proviso that the delta
must be understood according to the midpoint prescription when its argument
coincides with the upper boundary of integration as it occurs in practical
computations in perturbation theory around the white noise limit.

For any finite $\epsilon$ ordinary differential calculus holds true: the 
consistency conditions yields a Stratonovich discretisation prescription 
when $\epsilon$ is set to zero  and the recovery of the white noise advection
model of \cite{BBW}.

The information about the scaling of the correlations of the velocity field 
at equal times is stored in the constants $v_{m}$. We assume the power law
behaviour
\begin{equation}
|v_{m}|\propto k_{m}^{-\frac{\xi}{2}}.
\label{scalvel}
\end{equation}
Kolmogorov scaling is specified by $\xi=2/3$ while $\xi=2$ corresponds to
a laminar regime. 
The $\tau_{m}$'s in (\ref{ousde}) describe the typical correlation times
for the random velocity field. A simple physical interpretation is to identify
them with the turn-over times, i.e., with the typical time rates of variation 
through non linearity of the advection field on each shell
\cite{Vulpio}:
\begin{equation}
  \tau_{m} \sim \frac{1}{k_{m}|v_{m}|} \propto k_{m}^{-1+\frac{\xi}{2}}.
\label{tauscale} 
\end{equation}
The scaling of the correlation times is then fully specified in terms of
the parameter $\xi$. It is worth to note that for any $\xi$ less than two
the $\tau_{m}$'s are always decreasing functions of the wave number. 

The evolution of the scalar $\theta$ is determined in the inertial range
by its complex conjugate. It is useful to introduce an unified notation 
for the $2 N$ degrees of freedom. Calling $\Theta=\theta \bigoplus \theta^*$
 and $U=u \bigoplus u^*$ one has for the N shells:
\begin{equation}
\frac{d}{dt}\Theta_{\alpha}=\sum_{\beta=1}^{2N}[A_{\alpha,\beta}+
\sum_{\gamma=1}^{2N} B_{\alpha,\beta}^{\gamma}U_{\gamma}]\,\Theta_{\beta}
+f\delta_{\alpha,1}+f^*\delta_{\alpha,N+1}
\label{passcomplete}
\end{equation}
with
\begin{eqnarray}
&&A_{m,\beta}=-\kappa k_m^2 \delta_{m,\beta}\nonumber\\
&&A_{N+m,\beta}=-\kappa k_m^2 \delta_{m,n}\nonumber\\
&&B_{\alpha,\beta}^{m}=-ik_{m+1}[\delta_{\beta,m+1}\,\delta_{\alpha,N+m}-
\delta_{\beta,m}\,\delta_{\alpha,N+m+1}]\nonumber\\ 
&&B_{\alpha,\beta}^{N+m}=ik_{m+1}[\delta_{\beta,N+m+1}\,\delta_{\alpha,m}-
\delta_{\beta,N+m}\,\delta_{\alpha,m}]
\end{eqnarray}
where Latin and Greek indices range respectively from $1$ to $N$ and
from $1$ to $2 N$.  The set of matrices with constant entries
$B^{\gamma}$ do not commute within each other and with the $A$ matrix.
The known sufficient condition (see for example \cite{Arnold}) to have
a solution of (\ref{passcomplete}) in an analytic exponential form is
therefore not satisfied. From the geometrical point of view
non-commutativity means that the dynamics is confined on a manifold
which turns into an hyper-sphere in ${\cal C}^N$ in the inertial limit
(\ref{energy}).

The complex equations (\ref{passcomplete}) are invariant under phase
transformations. Given two diagonal hermitian $2N \times 2N$
matrices with time independent random entries
\begin{eqnarray}
&&T\equiv\mbox{diag}(e^{i\phi_1},...,e^{i\phi_N},e^{-i\phi_1},...,
e^{-i\phi_N})
\\
&&S\equiv\mbox{diag}(e^{-i(\phi_1+\phi_2)},...,e^{-i(\phi_{N-1}+\phi_N)},0,
e^{i(\phi_1+\phi_2)},...,e^{i(\phi_{N-1}+\phi_N)},0)
\end{eqnarray}
if $\Theta$ is a realisation of the solution of the equations of motion 
then
\begin{equation}
T \Theta(U)=\Theta(S U) 
\label{identity}
\end{equation}
is still a solution.  The phase symmetry is the remnant of the
translational invariance of the original hydrodynamical equations in
real space \cite{libro}.  From the phase symmetry (\ref{identity}) it
follows that at stationarity the only analytic non zero moments of the
correlation are of the form
\begin{equation}
C_{m_1,...,m_\omega}^{(2 \omega)}=\left<\Pi_{i=1}^{\omega}\Theta_{m_i}
\Theta_{N+m_i}\right>\equiv\left<\Pi_{i=1}^{\omega}|\theta_{m_i}|^2\right>.
\label{nonzero}
\end{equation}
In the inertial range such quantities display a power law behaviour.
The diagonal sector of the moments whose scaling properties are specified 
by the exponents $H(2\omega)$
\begin{equation}
C_{m,...,m}^{(2 \omega)} \propto k_m^{-H(2\omega)}
\label{Hdef}
\end{equation}
is in the shell model context the analogue of the structure
functions (\ref{structurefunc}) of the original PDE model (\ref{pde}).  
The exponents $H(2\omega)$'s are said to be normal if they can be derived 
from dimensional analysis. 
Under the assumption that a steady state is reached one matches the
scaling of the inertial terms in (\ref{passive}) with a power law decay 
of the solution
\begin{equation}
k_{m+1} k_m^{-\frac{\xi}{2}} \theta_{m+1}- k_{m} k_{m-1}^{-\frac{\xi}{2}} 
\theta_{m-1} \sim 0.
\label{matching}
\end{equation}
The resulting prediction is a linear behaviour of the exponents versus the 
order $\omega$ of the diagonal correlation:
\begin{equation}
H(2\omega)=\omega\,(1-\frac{\xi}{2}).
\label{normalscaling}
\end{equation}
The scaling argument (\ref{matching}) neglects completely the random
fluctuations of the passive scalar field. Normal scaling holds if the
statistics of the $\theta$-field is Gaussian. Deviations from normal
scaling are then related to the occurrence of intermittency
corrections to the Gaussian statistics. A systematic account of the
fluctuations is provided by the study of the equations of motion
satisfied by the moments of the scalar field.

\section{Equations of motion of the field moments}
\label{s:closure}

In the white noise limit, $\epsilon$ equal zero, the Furutsu-Donsker-Novikov 
formula \cite{Frisch} and the delta correlation in time of the velocity 
ensure that the moments $C^{(2 \omega)}$ are specified by the solutions 
of closed linear systems \cite{WB,BBW}. 
In the presence of finite time correlations stochastic calculus 
of variations \cite{Bass,Nualart} allows to write
non closed integro-differential equations for the correlations.
A typical functional integration by parts relation is: 
\begin{equation}
\left<F(\Theta(t))U_{N+m}(t)\right>=\int_{0}^{t}ds \left<U_{N+m}(t)U_m(s)
\right>\left<\frac{dF(\Theta(t))}{d \Theta_{\alpha}(t)}
R_{\alpha,\beta}(t,s)B_{\beta,\gamma}^{m}\Theta_{\gamma}(s)
\right>
\label{typical}
\end{equation}
where Einstein convention holds for repeated {\em Greek} indices. The 
matrix $R$ is the fundamental solution of the homogeneous system 
associated with (\ref{passcomplete}).
A heuristic proof of the stochastic integration by parts formula and of 
(\ref{typical}) is provided in Appendices \ref{ap:one} and \ref{ap:two}. 

Let us start with the second moment of the scalar field
\begin{equation}
C_m^{(2)}(t)\dot{=}\left<\Theta_{m}(t)\Theta_{N+m}(t)\right>\equiv
\left<\theta_{m}(t)\theta_{m}^*(t)\right>.
\end{equation}
>From the equations of motion (\ref{passcomplete}) one has
\begin{eqnarray}
\lefteqn{[\frac{d}{dt}+2 \kappa k_m^2\,] C_m^{(2)}(t)
-2 \Re\left\{\left<\Theta_{N+m}(t)f(t)\right>\right\}\delta_{m,1}=}\nonumber\\
&&+2 \Re\left\{i\,k_{m+1}\left<U_{N+m}(t)\Theta_{N+m+1}(t)\Theta_{N+m}(t)
\right>\right\}+\nonumber\\
&&-2\,k_{m}\Re\left\{i\,\left<U_{N+m-1}(t)\Theta_{N+m-1}(t)\Theta_{N+m-1}(t)
\right>\right\}.
\label{2pointsdiff}
\end{eqnarray}
The integration by parts formula (\ref{typical}) gives 
\begin{eqnarray}
\lefteqn{[\frac{d}{dt}+2 \kappa k_m^2] C_m^{(2)}(t)-2 \delta_{m,1}\Re\,
\int_{0}^{t} \,ds \,\left<f(t)f(s)^*\right>\left<R_{N+m,N+1}(t,s)\right>=}
\nonumber\\
&&2\,k_{m+1}^2 \tau_{m} \int_{0}^{t}\,ds \,
\frac{\left<U_{m}(t)U_{N+m}(s)\right>}{\tau_m}
\Re{\cal F}_m^{(2)}(t,s)+\nonumber\\
&&-2\,k_{m}^2 \tau_{m-1}  \int_{0}^{t}\,ds \,
\frac{\left<U_{m-1}(t)U_{N+m-1}(s)\right>}{\tau_{m-1}}
\Re{\cal F}_{m-1}^{(2)}(t,s)
\label{2points}
\end{eqnarray}
where $m=1, \ldots , N$, $\Re$ is the real part and:
\begin{eqnarray}
\label{Ffunc}
&&{\cal F}_m^{(2)}(t,s)\dot{=} G_{N+m+1,N+m\,;\,N+m,m+1}^{(2)}(t,s)-
G_{N+m+1,N+m+1\,;\,N+m,m}^{(2)}(t,s)\nonumber\\
&&+G_{N+m,N+m\,;\,N+m+1,m+1}^{(2)}(t,s)-G_{N+m,N+m+1\,;\,N+m+1,m}^{(2)}(t,s)\\
\label{Gfunc}
&&G_{N+m,N+n\,;\,N+p,q}^{(2)}(t,s)\dot{=}\sum_{\alpha=1}^{2N}
\left<\Theta_{N+p}(t) R_{N+m+1,\alpha}(t,0)R_{\alpha,N+n}^{-1}(s,0)
\Theta_{q}(s)\right>\\
\label{dvalues}
&&d_{m}\dot{=}|v_{m}|^{2}\,\tau_m \propto k_{m}^{-(1+\frac{\xi}{2})}
\end{eqnarray}
When a steady state is reached the LHS of (\ref{2points}) can be neglected
through the whole inertial range. The RHS specifies the inertial
operator of the theory.
A further simplification is attained in the limit of very large shell number.
For any $\xi$ less than two one has
\begin{eqnarray}
 \lim_{m \uparrow \infty} \frac{\left<U_{m}(t)U_{N+m}(s)\right>}{\tau_m}
\equiv
\lim_{m \uparrow \infty} \frac{\left<u_{m}(t)u_{m}^{*}(s)\right>}{\tau_m}=
|v_{m}|^{2}\,\delta(t-s)
 \label{largemcorr}
\end{eqnarray} 
independently of $\epsilon$. 
At equal times the resolvent matrix $R$ reduces to the identity.
>From (\ref{scalvel}) and (\ref{tauscale}) it stems that
\begin{equation}
 k_{m+1}^2 d_{m}\,\tau_{m}= \lambda^2.
 \label{coeff}
\end{equation}
Hence for $m$ going to infinity the inertial operator is linearised in 
the form
\begin{eqnarray}
RHS=2 \frac{\lambda^2}{\tau_m}\left(C_{m+1}^{(2)}-C_{m}^{(2)}\right)-
2\,\frac{\lambda^2}{\tau_{m-1}}\left(C_{m}^{(2)}-C_{m-1}^{(2)}\right).
\label{largem}
\end{eqnarray}
The slowest decay scaling solution compatible with a zero LHS is 
\begin{equation}
 C_{m}^{(2)}\propto \tau_m=k_m^{-H(2)},
 \label{2pointnorm}
\end{equation}
In other words we have proven that the scaling of the second moment is 
normal as it coincides with the dimensional prediction (\ref{normalscaling}). 
Moreover since the result does not depend on $\epsilon$ it is universal 
versus the time correlation. 
It is worth stressing that the derivation of (\ref{2pointnorm}) requires that 
each of the terms appearing in (\ref{largem}) has separately a finite 
non zero limit for $m$ going to infinity.  
The condition turns out to be not self-consistent when the same reasoning
is applied to moments higher than the second. 
  
An important consequence of normal scaling of the $C_{m}^{(2)}$'s is the
Obukhov-Corrsin \cite{Obukhov,Corrsin} law for the decay of the power 
spectrum $\Gamma(k)$ of the passive scalar if the Kolmogorov scaling is 
assumed for the advecting field:
\begin{equation}
  \Gamma(k)=
  \frac{d}{dk}\sum_{k_{n} \leq k} \left<(\theta_n \theta_n^*)^2\right> 
  \propto k^{-(H(2)+1)}|_{\xi=2/3}=k^{-\frac{5}{3}}.
\label{OC}
\end{equation}

A second  interesting limit is when $\epsilon$ tends to zero. Neglecting
all non stationary contributions to the velocity correlations the RHS
of (\ref{2points}) becomes
\begin{eqnarray}
\lefteqn{RHS=2\,k_{m+1}^2 d_{m} \left(C_{m+1}^{(2)}-C_{m}^{(2)}\right)-
2\,k_{m}^2 d_{m-1}\left(C_{m}^{(2)}-C_{m-1}^{(2)}\right)+}\nonumber\\
&&-2k_{m+1}^2 d_{m} \int_{0}^{t} \,ds e^{-\frac{t-s}
{\epsilon \tau_m}}\frac{d}{ds}\Re{\cal F}_m^{(2)}(t,s)+
\nonumber\\
&&+k_{m}^2 d_{m-1}\int_{0}^{t}\,ds\,e^{-\frac{t-s}{\epsilon \tau_{m-1}}} 
\frac{d}{ds}\Re{\cal F}_{m-1}^{(2)}(t,s).
\label{2pointbyparts}
\end{eqnarray}
If $\epsilon$ is set exactly to zero the integral terms disappear
and the white noise equations of \cite{BBW} are recovered. 
The information about the scaling of the velocity field is absorbed
in the $d_m$'s. In a pure white noise theory it is convenient
to redefine the turbulence parameter as
\begin{equation}
  \xi_{wn} = 1 + \frac{\xi}{2}.
\end{equation}
A Kolmogorov scaling of the velocity field corresponds to $\xi_{wn}$ equal 
to four thirds which is also the value giving the Obukhov-Corrsin scaling 
in (\ref{OC}). The two definitions of the degree of turbulence coincide 
for $\xi$ equal to two (Batchelor limit). It is natural to identify 
$\xi_{wn}$ with the turbulence parameter of the Kraichnan model. 
The correspondence fixes the physical range of $\xi$ between $[-2,2]$.

In the general case of the $2\,\omega$-th even moment of the scalar 
$C^{(2\,\omega)}$  one has:
\begin{eqnarray}
&&RHS=\sum_{q_{1},...,q_{\omega}}\,
I_{m_{1},...,m_{\omega},q_{1},...,q_{\omega}}^{(2\,\omega;0)}\,C^{(2\,\omega)}_
{q_{1},...,q_{\omega}}+\nonumber\\
&& - \sum_{i=1}^{\omega}\, 2k_{m_{i}+1}^2 d_{m_{i}}\,
\int_{0}^{t}\,
e^{-\frac{t-s}{\epsilon \tau_{m_{i}}}}\frac{d}{ds}
\Re{\cal F}_{m_{1},...,m_{i},...,m_{\omega}}^{(2\,\omega)}(t,s)+
\nonumber\\
&&+\sum_{i=1}^{\omega}\,2 k_{m_{i}}^2 d_{m_{i}-1}\,
\int_{0}^{t}ds\,
e^{-\frac{t-s}{\epsilon \tau_{m_{i}-1}}} \frac{d}{ds}
\Re{\cal F}_{m_{1},...,m_{i}-1,...,m_{\omega}}^{(2\,\omega)}(t,s).
\label{2omegabyparts}
\end{eqnarray} 
The multidimensional matrix $I^{(2\,\omega;0)}$ is the linear inertial 
operator of the white noise theory.
The integrand functions ${\cal F}_{m_{1},...,m_{i}-1,...,
m_{\omega}}^{(2\,\omega)}(t,s)$ are given by the straightforward generalisation
of (\ref{Ffunc}). The LHS, as above, is set to zero as far as the 
steady state features of the inertial range are concerned.
Repeated integrations by parts in the large $t$ limit generate a 
Laplace asymptotic expansion \cite{Erdelyi} of integral terms in the RHS, 
the coefficient of which are the 
derivatives with respect to $s$ of the functions ${\cal F}^{(2 \omega)}$ 
evaluated at equal times. When the steady state sets in we assume
the latter quantities to be invariant under time translations for large $t$. 
Under such an assumption it will be proven in section~\ref{s:perturbative} 
that the equal time derivatives are specified at equilibrium by linear 
combinations of the $C^{(2 \omega)}$'s.
The effect of a small time correlation is therefore to generate new couplings 
of order $\epsilon$ in the inertial operators. 
The observables we focus on are the scaling exponents. As discussed in 
the introduction, anomalies occur in the presence of non trivial scaling zero 
modes of the white noise inertial operators. It makes sense to relate the 
$\epsilon$ dependence of the anomalous exponents to a perturbation of the 
scaling zero modes derived for $\epsilon$ equal zero. A straightforward 
approach to the problem calls for the solution of $N^{\omega}$ linear 
equations. 
A further source of difficulty is that the exact determination of the zero
eigenvectors of the inertial operators of any given order requires the matching
of infra-red and ultra-violet boundary conditions. In the absence of 
an exact diagonalisation any analytical approach must rely on  closure 
Ans\"atze to solve first the white noise problem and then to yield the 
corrections to the zero modes by linear perturbation theory.

\section{The white noise closure}
\label{s:wnclosure}
\begin{figure}[htbp]
  \begin{center} 
  \epsfig{file=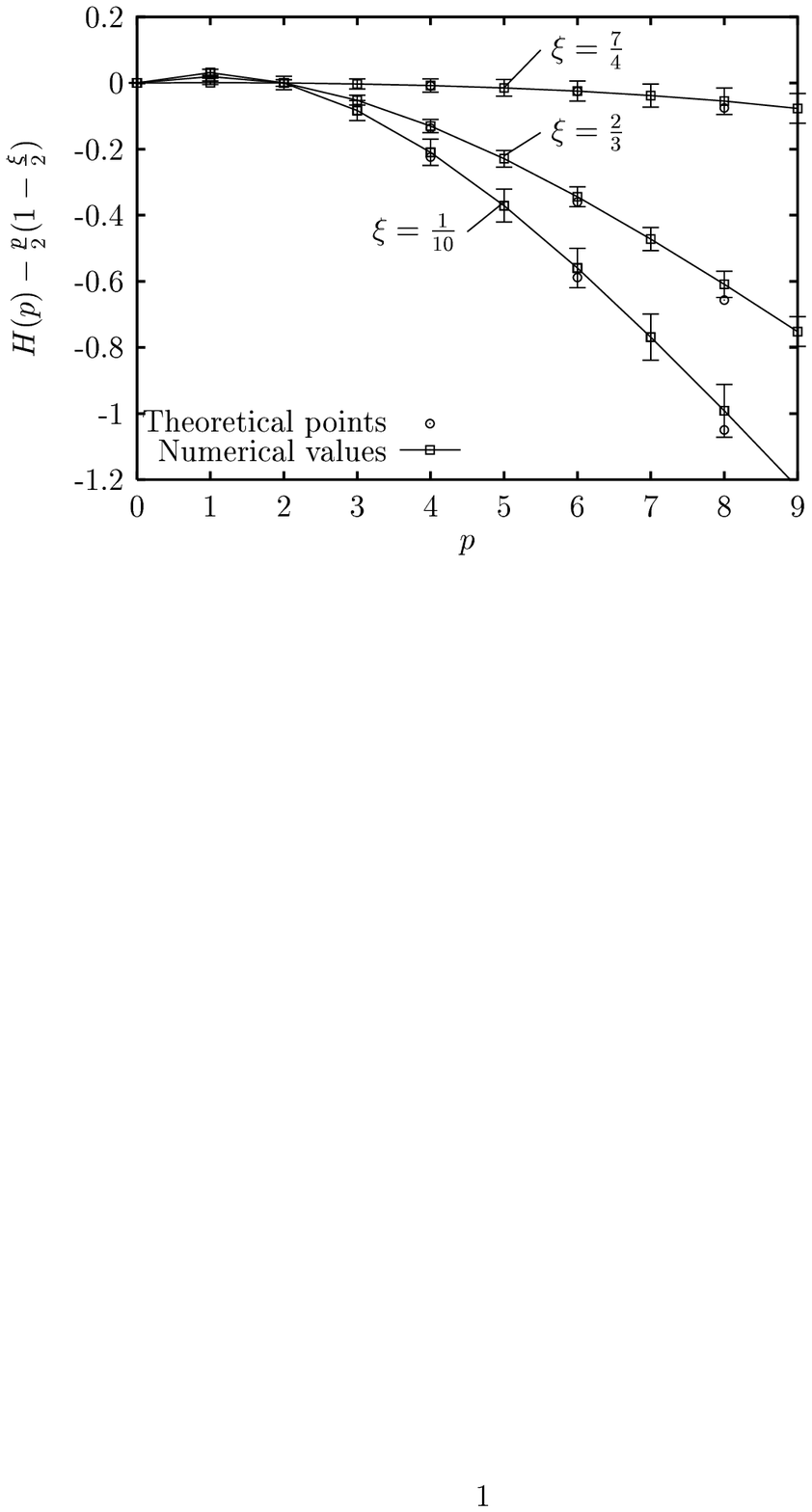} 
  \caption{The analytical
    prediction for the anomalous part of the scaling exponent compared
    with the result of the numerical experiments for different values
    of the turbulence degree parameter $\xi$.  A Kolmogorov scaling of
    the advection field corresponds to $\xi=2/3$.  The dash-dotted
    line represents the (dimensional) normal scaling prediction.  The
    continuum line interpolates the exponents as obtained from the
    numerical experiment (squares). The circles are the analytical
    prediction from the closure Ansatz.}  
  \label{fig:wn} 
  \end{center}
\end{figure}

In the present section we present a closure strategy to compute the
$H(2 \omega)$'s in the case of white noise advection. 
As shown in the previous section the second diagonal moment is normal and 
universal versus the time correlation. 
The first nontrivial zero mode problem is provided by the fourth order 
inertial operator
$I^{(4;0)}$.
In \cite{BBW} it has been shown that the anomalous exponent 
$\rho_4$
\begin{equation}
H(4)=2 H(2)-\rho_4
\end{equation}
can be extracted up to a very good accuracy from the solution of only
two non-linear algebraic equations. The stationary equations for 
$C^{(4)}$ in the inertial range far from the infra-red and ultra-violet 
boundaries are given by
\begin{eqnarray}
\lefteqn{0=\frac{I_{m,n;p,q}^{(4;0)}C_{p,q}^{(4)}}{2\,\lambda^2}
\equiv }\nonumber\\
&&\equiv -(\frac{1}{\tau_m}+\frac{1}{\tau_{m-1}}+\frac{1}{\tau_n}
+\frac{1}{\tau_{n-1}}) C_{m,n}^{(4)}+\frac{1}{\tau_m} C_{m+1,n}^{(4)}+
\frac{1}{\tau_n} C_{m,n+1}^{(4)}+\nonumber\\
&&+\frac{1}{\tau_{m-1}} C_{m-1,n}^{(4)}+\frac{1}{\tau_{n-1}} C_{m,n-1}^{(4)}
+2\, \delta_{m,n}(\frac{C_{m,m+1}^{(4)}}{\tau_{m}}+ 
\frac{C_{m,m-1}^{(4)}}{\tau_{m-1}})+ 
\nonumber\\
&&-2\,\delta_{m+1,n}\,\frac{ C_{m,m+1}^{(4)}}{\tau_m}-2\,
\delta_{n,m-1}\,\frac{ C_{m,m-1}^{(4)}}{\tau_{m-1}}.
\label{irwn4}
\end{eqnarray}
One recognises two kinds of couplings in $I_{m,n;p,q}^{(4;0)}$. 
\begin{itemize}
\item ``Global'' or ``unconstrained'' interactions: the indices $p$
and $q$ range respectively from $m-1$ to $m+1$ and from $n-1$ to
$n+1$. The coupling are independent upon the relative values of $m$
and $n$: in this sense they are referred as global.
\item ``Purely local'' interactions: they occur only for $|m-n| \leq
1$ and correspond to the terms proportional to the Kroenecker's
$\delta$ in (\ref{irwn4}).
\end{itemize}
Anomalous scaling in the inertial range is strictly related to the presence 
of such purely local interactions. Were these latter neglected the fourth 
order moment would have a normal scaling solution
\begin{equation}
 C_{m,n}^{(4)} \propto \frac{\tau_n}{\tau_m} \tau_m^2.
\label{4normal}
\end{equation}
The idea is to capture the anomalous scaling by looking at the 
``renormalisation'' of global couplings by pure short range ones.
Disregarding the boundaries, the system is invariant under a simultaneous 
shift of the indices. Hence assuming a perfect index-shift invariance 
there are, for the $m$-th shell, only two independent equations where 
$\delta$-like terms occur:
\begin{eqnarray}
&&0=\sum_{p,q}\,I_{m,m;p,q}^{(4;0)}C_{p,q}^{(4)} 
\nonumber\\
&&0=\sum_{p,q}\,I_{m,m-1;p,q}^{(4;0)}C_{p,q}^{(4)}.
\label{4closed}
\end{eqnarray}
The third equation involving a purely local interaction of the $m$-th 
shell with its nearest neighbours
\begin{eqnarray}
0=\sum_{p,q}\,I_{m+1,m;p,q}^{(4;0)}C_{p,q}^{(4)} \nonumber.
\end{eqnarray}
is generated from the second of the (\ref{4closed})'s by a simple index shift.
Therefore it is not regarded as independent.
The pair (\ref{4closed}) contains all the relevant information
to extract the scaling of the fourth moment. It forms a closed system of
equations independently on the shell number $m$ as one imposes scaling 
relations to hold within the set of ``independent'' moments of fourth order: 
\begin{eqnarray}
&&C_{m+n,m+n}^{(4)}=z^{-n}C_{m,m}^{(4)} 
\label{4diagonal} \\
&&C_{m+n,m}^{(4)}=x\,k^{-H(2)}_{n-1}\,C_{m,m}^{(4)}
 \label{4nondiagonal}
\end{eqnarray}
where the integer $n$ is taken larger than zero.
As in the analysis of the interactions the concept of independence stems
from the assumption of index shift invariance: the moments of the form 
$C_{m-n,m}^{(4)}$ are immediately reconstructed once (\ref{4diagonal}) and
(\ref{4nondiagonal}) are given:
\begin{eqnarray}
C_{m-n,m}^{(4)}=x\,k^{-H(2)}_{n-1}\,z^{n}C_{m,m}^{(4)} \nonumber
\end{eqnarray}
Let us analyse the closure Ansatz in more detail. The first equation
(\ref{4diagonal}) is a global scaling assumption of the ``diagonal''
sector of the fourth moment. Its justification lies in the very
definition of an inertial range. The second scaling assumption relates
the diagonal sector to the non diagonal one via a marginal scaling. It is
the analogous in the present context of an operator product expansion
(OPE) in statistical field theory \cite{Cardy}. There renormalisation
group (RG) techniques are able to describe the scaling behaviour of
correlations of fields sampled at large real space distances one from
the other. If an observable requires the evaluation of a correlation
including the products of one field in two points at short distance
i.e. $\left<\phi(x-dx) \phi(x+dx)...\right>$ the RG procedure cannot
be directly applied.  The problem is overcome by an OPE or short
distance expansion. The prescription is to rewrite the product via a
Taylor expansion in terms of local composite operators sampled just in
one point. Such a point is now well separated from all the other
appearing in the correlation function.  The original correlation is
substituted by a set correlations such that RG applies provided an
extra renormalisation, renormalisation of composite operators (RCO),
is introduced.  The latter is understood by observing that in our
example the first term in the Taylor expansion gives
\begin{eqnarray}
\phi(x+dx) \phi(x-dx) \sim  \phi(x)^{2} \nonumber
\end{eqnarray}
The mathematical meaning of a field is that one of an operator-valued 
distribution. The product of two distributions at equal points i.e. 
$\phi(x)^{2}$ requires a regularisation before the cut-off is removed in 
order to make itself sense as a distribution. This is the content of the RCO.
Finally at leading order the relation between the renormalised quantities reads
for the above example
\begin{eqnarray}
\left< \left[\phi(x+dx) \phi(x-dx)\right]_{R}...\right> \sim c(dx)\left<
\left[\phi(x)^{2}\right]_{R}...\right>.
\label{OPE}
\end{eqnarray}
Roughly speaking the small real space separations are associated with the
UV behaviour of the Fourier conjugated variable.
In the shell model context the $\theta_{m}$ are representative of the
scalar field variation over one octave. The moments $C_{m,m+n}^{(4)}$  
correspond to the Fourier transform of the fourth order structure functions 
of a homogeneous and isotropic real space solution of the passive scalar 
equation (\ref{pde}).
The meaning of the non-diagonal closure (\ref{4nondiagonal}) is to assume
the long range (many shells) behaviour of the Fourier transform of the
OPE coefficient $c(dx)$ to scale inside the inertial range for large $n$ 
independently on $m$. The constant $x$ renormalises the value of the 
first shell where deviation from scaling occur. The analogy with the
OPE is then summarised by writing
\begin{eqnarray}
&&\phi(x)=[\theta(x)-\theta(0)]^{2}_{R} \nonumber\\
&&\lim_{|k|\uparrow \infty} \int d^{D}x e^{i k \cdot dx} c(dx) \sim 
x\,k_{n-1}^{-H(2)}
\end{eqnarray}
The insertion of the scaling Ansatz in (\ref{4closed}) leaves a
non-linear system in the unknown variables $z$ and $x$.
By applying the definition $k_{n}=\lambda^{n}$ one gets
\begin{equation}
  \left\{%
    \begin{array}{l}
      -1 - {{\lambda }^{-H(2)}} + 2\,x\,(1 + z\,{{\lambda
          }^{-H(2)}})=0 \nonumber\\ 
      ( 1 + z ) \,{{\lambda }^{-H(2)}} + x\,z\,( -1 - 3\,{{\lambda
          }^{-H(2)}} {{\lambda }^{-2\,H(2) }} + z\,{{\lambda
          }^{-3\,H(2) }})=0 
    \end{array}
  \right.
\end{equation}
which after a straightforward manipulation provides $z$ as the physical root
of a second order polynomial
\begin{eqnarray}
&&z=\frac{1+2\ {{\lambda }^{-H(2) }}+
2\ {{\lambda }^{-2 H(2) }}+{{\lambda }^{-3 H(2) }}+
\sqrt{1+4\ {{\lambda }^{- H(2) }}+8\ {{\lambda }^{- 2 H(2) }}-6\ 
{{\lambda }^{- 3 H(2) }}-4\ {{\lambda }^{- 5 H(2) }}+
{{\lambda }^{- 6 H(2) }}}}{2\ (2\,{\lambda }^{- 2 H(2) }+
{{\lambda }^{- 3 H(2) }}+{{\lambda }^{- 4 H(2) }})}.  
\end{eqnarray}  
In terms of $z$ the anomaly is
\begin{eqnarray}
\rho_4=2\,H(2)-\frac{\log{z}}{\log{\lambda}}
\end{eqnarray} 
and it proves proves to be in fair agreement with the values obtained
from the numerical solution of the exact equations (\ref{irwn4})
\cite{BBW} and from the numerical integration of (\ref{passive}) for
all the values of the turbulent exponent $\xi$ in the physical range
$[0,2]$ (see also Fig.~\ref{fig:wn}). The sign of $\rho_4$ is always
positive: the effect of the anomaly is to decrease the diagonal
scaling exponent.

The procedure presented in detail for the computation of the
fourth order exponent is straightforwardly extended to any higher 
order moment when one recognises that two crucial observation hold 
in general.
\begin{itemize}
\item In the absence of pure short range couplings the normal
scaling prediction holds true far from the boundaries for the
zero modes of the inertial operators of any order $2 \omega$. 
\item For any fixed shell $m$ there is a one to one correspondence 
between the number of independent equations and moments of order $2 \omega$.
\end{itemize} 
In the case of $C^{(2 \omega)}$ there are $2^{\omega-1}$ equations: 
for any fixed reference shell $m_1$ the interaction with the second 
index $m_2$ is affected by a pure short range coupling if the latter 
is equal or one unit different from $m_1$ i.e. only two possible choices 
and so on until the $\omega$-th index is reached. On the other hand 
$2^{\omega-1}$ is the number of exponents which characterise the scaling 
of the $2 \omega$-th moment. The Ansatz is that the marginal
scaling of the non-diagonal sector is fully specified in terms of the
diagonal scaling exponents of order less than $2\,\omega$. By means
of the ``OPE's'' one is able to close the zero mode equations in terms 
of $2^{\omega-1}$ unknown renormalisation constants and $H(2 \omega)$. 
The analogy with a field theoretical OPE goes to show that the need for 
an infinite set of constant does not necessarily imply the 
non-renormalisability of the real space theory mimicked by the shell 
model \cite{Zinn}.

More concretely the diagonal scaling exponent of the sixth moment
($\omega=3$) of the scalar field
\begin{equation}
C_{m,n,p}^{(6)}=\left<\Theta_{N+m} \Theta_m \Theta_{N+n} \Theta_n
\Theta_{N+p} \Theta_p \right> \equiv
\left<|\theta_{m}|^{2}|\theta_{n}|^{2}|\theta_{p}|^{2}\right>
\label{6nondiagonal}
\end{equation}
according to the above criterion requires four independent equations 
(see appendix \ref{ap:sixthorder})
\begin{equation}
  \left\{%
    \begin{array}{l}
\sum_{p,q,r}\,I_{m,m,m;p,q,r}^{(6;0)}C_{p,q,r}^{(6)}=0 \nonumber\\
\sum_{p,q,r}\,I_{m,m,m-1;p,q,r}^{(6;0)}C_{p,q,r}^{(6)}=0 \nonumber\\
\sum_{p,q,r}\,I_{m,m-1,m-1;p,q,r}^{(6;0)}C_{p,q,r}^{(6)}=0 \nonumber\\
\sum_{p,q,r}\,I_{m,m+1,m-1;p,q,r}^{(6;0)}C_{p,q,r}^{(6)}=0 
\label{6closed}
    \end{array}
  \right.
\end{equation}
The OPE inspired closure yields
\begin{eqnarray}
C_{m+n,m+n,m+n}^{(6)}&&= z^{-l}C_{m,m,m}^{(6)}\nonumber\\
C_{m+n,m+n,m}^{(6)}&&= x_{1}\,k^{-H(4)}_{n-1}  C_{m,m,m}^{(6)}
\nonumber\\
C_{m+n+p,m+n,m}^{(6)}&&=x_{2}\,k^{-H(2)}_{p-1} k^{-H(4)}_{n-1} 
C_{m,m,m}^{(6)}\nonumber\\
C_{m+n,m,m}^{(6)}&&=x_{3}\,k^{-H(2)}_{n-1} C_{m,m,m}^{(6)}
\label{6scaling}
\end{eqnarray}
Inserting the ``OPE'' in (\ref{6closed}) one gets into the algebraic system
for the unknown renormalisation constants $(x_1,x_2,x_3)$ and the diagonal
scaling factor $z$.
\begin{eqnarray}
&&-1 + {{\lambda }^{\alpha }}\,
   \left( -1 + 3\,z\,{x_1} \right)  + 3\,{x_3}=0\nonumber\\
&&-{{\lambda }^{2\,\alpha }}\,z\,{x_1}   + 
  {{\lambda }^{4\,\alpha  + {{\rho }_4}}}\,{{z}^2}\,
   {x_1} - 2\,z\,\left( {x_1} - 2\,{x_2} \right)  + 
  {{\lambda }^{\alpha }}\,
   \left( 1 + z\,\left( -7\,{x_1} + 4\,{x_3} \right)
         \right)=0\nonumber\\
&&{{\lambda }^{\alpha }}\,\left( 1 + 4\,{x_1} - 
     6\,{x_3} \right)  + 
  {{\lambda }^{2\,\alpha }}\,
   \left( 4\,z\,{x_2} - 2\,{x_3} \right)  - {x_3}=0
\nonumber\\
&&{{\lambda }^{3\,\alpha  + {{\rho }_4}}}\,z\,{x_1} + 
  {{\lambda }^{\alpha }}\,z\,
   \left( {x_1} - 3\,{x_2} \right)  - z\,{x_2} + 
  {{\lambda }^{5\,\alpha  + {{\rho }_4}}}\,{{z}^2}\,
   {x_2} - {{\lambda }^{3\,\alpha }}\,z\,
   \left( {x_2} - {x_3} \right)  + 
  {{\lambda }^{2\,\alpha }}\,
   \left( -4\,z\,{x_2} + {x_3} \right)=0
\label{6nonlin}
\end{eqnarray}
After some algebra (\ref{6nonlin}) reduces to a single third order polynomial 
specifying the physical root of $z$. It is worth to remark that from the
functional dependence of the coefficient of (\ref{6nonlin}) the exponent
$H(6)$ depends upon the anomaly of $H(4)$. Once again the anomaly 
evaluated from 
\begin{equation}
\rho_6=3\,H(2)-\frac{\log{z}}{\log{\lambda}}
\end{equation} 
is in fair agreement with numerics (see Fig.~\ref{fig:wn}) for 
different values of $\xi$.

\begin{figure}
 \begin{center} 
  \epsfig{file=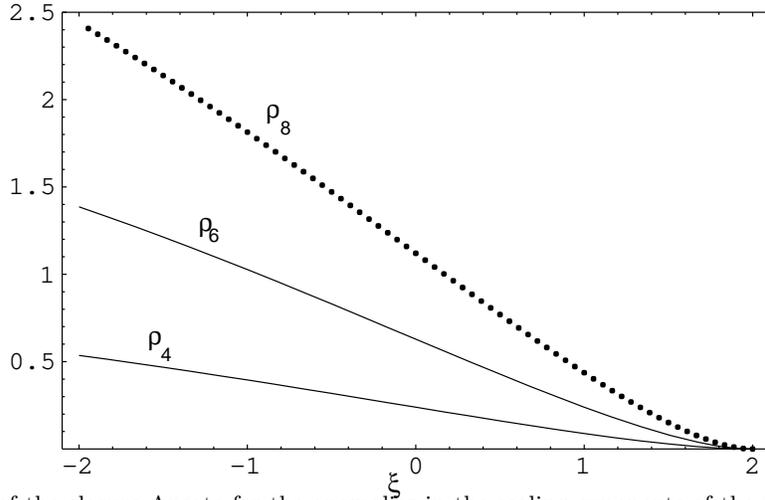} 
  \caption{The prediction of
    the closure Ansatz for the anomalies in the scaling exponents of
    the fourth, $\rho_4$, the sixth, $\rho_6$ and the eight, $\rho_8$,
    moments of the scalar field versus the turbulence parameter $\xi$.
    In all cases the anomalies are decreasing function of $\xi$ going
    smoothly to zero as the Batchelor limit $\xi$ equal to two is
    approached. The anomaly of the eight moment is obtained as the
    numerical solution of a sixth order polynomial.}
  \label{fig:wnanomal}
 \end{center}
\end{figure}

In appendix~\ref{ap:eightorder} the same steps are performed in the
case of the eight moment $C_{m,n,p,q}^{(8)}$. The analytical
predictions for the anomalous exponents are summarised in
Fig.~\ref{fig:wnanomal}.  In all cases the anomalies are decreasing
functions of the turbulence parameter $\xi$ vanishing smoothly when
the laminar limit ($\xi$ equal two) is approached.  The anomaly of the
fourth order moment can be compared with the results of numerical
experiments for the fourth order structure function of the Kraichnan
model \cite{FMV}.  There the adopted turbulence parameter is
$\xi_{wn}$. For values of $\xi_{wn}$ of order one, i.e., from the
Kolmogorov scaling up to the Batchelor limit one indeed observes the
same monotonically decreasing behaviour with values of the anomaly of
the same order of those found in the shell model.  For lower values of
$\xi_{wn}$ the anomaly in the Kraichnan model display a maximum before
decreasing to zero for $\xi_{wn}$ equal to zero, i.e., when $\xi$
tends to minus two.  No sign of such behaviour is observed in shell
model.  The discrepancy might be an artifact of the shell model, which
was originally designed to mimic the supposed local in scale character of
the nonlinear interactions in a turbulent flow \cite{libro},
fails to describe a r\'egime where strong non local effects become
important.

On a phenomenological level the energy transfer in the inertial range of 
turbulent field is related to the occurrence of a cascade mechanism as 
firstly conjectured by Richardson \cite{Richardson}.
The conservation of energy in the inertial range imposes that the forcing
occurring on large real space scales is transferred to small scales 
(i.e. large wave numbers) before being dissipated.
A mathematical description of the cascade is provided by multiplicative 
stochastic processes \cite{Parisi}. Multiplicative modelling has been shown 
to account for most of the features observed in real and synthetic 
turbulence \cite{multi,libro}.
In the present case the idea of a multiplicative structure is incorporated
in the hypothesis that the scaling of the non diagonal sector of a given
moment of order $2\,\omega$ is reconstructed once the scaling of the lower 
moments is known. Such an assumption together with the analysis of the 
couplings in the inertial operator of order $2\,\omega$ 
yields with fair accuracy the scaling exponents of the model without the 
resort to an exact diagonalisation of inertial operator.

\section{Perturbative analysis}
\label{s:perturbative}
Let us now turn to the time correlated case. The idea is to evaluate 
the scaling behaviour of the dominant zero modes of the inertial operators  
(\ref{2omegabyparts}) linearised up to first order in $\epsilon$ 
by perturbing the white noise closure Ansatz.

The first order corrections in $\epsilon$ to the inertial operators 
are obtained by truncating the integration by parts to the terms linear
in $\epsilon$
\begin{eqnarray}
\lefteqn{2k_{m_{i}+1}^2 d_{m_{i}}\,\int_{0}^{t}\,
e^{-\frac{t-s}{\epsilon \tau_{m_{i}}}}\frac{d}{ds}
\Re{\cal F}_{m_{1},...,m_{i},...,m_{\omega}}^{(2\,\omega)}(t,s)=}\nonumber\\
&&=2\,\epsilon \,\lambda^2 \,\frac{d}{ds}
\Re{\cal F}_{m_{1},...,m_{i},...,m_{\omega}}^{(2\,\omega)}(t,s)|_{s=t}+
O(\epsilon^2,\epsilon\,e^{-\frac{t}{\epsilon \tau_{m_{i}}}}).
\label{correctiongen}
\end{eqnarray}
The use of (\ref{coeff}) in the RHS stresses that the effective
a-dimensional expansion parameter is $\epsilon \lambda^2$: the range of 
reliability of first order perturbation theory is compressed to 
$\epsilon \leq \lambda^{-2}/10$.
As mentioned in section~\ref{s:closure}, in the limit $t$ going to infinity,
one expects two time quantities to be stationary. In such a case the 
derivative with respect to the variable $s$ can be interchanged with the 
derivative with respect to $t$ and one can use the equations of motion to 
evaluate (\ref{correctiongen}). A direct differentiation with respect 
to $s$ is consistently taken with respect to the system of stochastic 
differential equations conjugated by time reversal of equations 
(\ref{passive}) to (\ref{force}). 
The latter operation in general requires the knowledge of the probability 
density of the forward problem. In the stationary limit the time reversal 
operation for the O-U process reduces to the inversion of the sign of the 
drift term as in the deterministic case. After a slightly more lengthy 
algebra the result is equal to the differentiation with respect to $t$ 
with opposite sign.     

The computations in the general case are very cumbersome (see
appendices~\ref{ap:fourthorder}, \ref{ap:sixthorder} and
\ref{ap:eightorder}).  It is convenient to exemplify the procedure in
the simpler case of the second order correlation.  There are four
contributions to $\Re{\cal F}_{m}^{(2)}$:
\begin{eqnarray}
&&\frac{d}{dt}G_{N+m+1,N+m\,;\,N+m,m+1}^{(2)}(t,s)|_{t=s}=0\nonumber\\
&&\frac{d}{dt}G_{N+m+1,N+m+1\,;\,N+m,m}^{(2)}(t,s)|_{t=s}=
-\kappa k_{m+1}^{2}C_{m}^{(2)}(t)+
\left<\dot{\Theta}_{N+m}(t)\Theta_{m}(t)\right>\nonumber\\
&&\frac{d}{dt}G_{N+m,N+m+1\,;\,N+m,m+1}^{(2)}(t,s)|_{t=s}=0\nonumber\\
&&\frac{d}{dt}G_{N+m,N+m\,;\,N+m+1,m+1}^{(2)}(t,s)|_{t=s}=
-\kappa k_{m}^{2}C_{m}^{(2)}(t)+\left<\dot{\Theta}_{N+m+1}(t)
\Theta_{m+1}(t)\right>.
\end{eqnarray}
By definition
\begin{eqnarray}
&&\left<\dot{\Theta}_{N+m}(t)\Theta_{m}(t)\right>=
\frac{1}{2}\frac{d}{dt}\left<|\theta_{m}(t)|^{2}\right>
\nonumber\\
&&
\left<\dot{\Theta}_{N+m+1}(t)
\Theta_{m+1}(t)\right>=\frac{1}{2}\frac{d}{dt}
\left<|\theta_{m+1}(t)|^{2}\right>.
\nonumber
\end{eqnarray}
The time derivative of $\Re{\cal F}_{m-1}^{(2)}$ is derived by a simple 
index shift. As the terms non-diagonal in the resolvent $R$ indices
are zero, the second moment inertial operator is not by affected by
first order perturbation theory. The result is not surprising.
The second moment has only one free index. Hence at any order of
perturbation theory only global coupling can be generated which are
forced by the symmetries of the model to be consistent with a normal
scaling of the zero mode. Moreover the $\Theta_{m}$'s components of
the scalar evolve only through the coupling with their complex conjugated
$\Theta_{N+m}$'s: their variation is a second order effect in $\epsilon$.
The only possible non zero corrections are viscous and can be
consistently neglected.

Let us turn to draw the general picture when $\omega$ is larger than
one.  Once again the phase symmetries (\ref{identity}) and the fact that
when $\Re{\cal F}_{m_{1},\ldots,m_{i},\ldots,m_{\omega}}^{(2\,\omega)}$ is
known all other terms are yielded by index shift or exchange
operations, prevent the corrections to the global couplings from
affecting the scaling properties: the resulting ``global'' sectors of
inertial operators have a normal scaling zero mode. This is in
agreement with the observation made in \cite{Chertkov}, where the
dependence of the scaling exponents on the time correlation for
generalised models of passive scalar advection is predicted to appear
only through anomalies. The corrections to the purely short range
couplings are therefore the relevant ones. They occur in two ways. On
one hand new terms of order $\epsilon$ show up in the purely self and
nearest-neighbours interactions.  On the other hand, terms proportional to
$\delta_{m_{i},m_{j}\pm 2}$ appear.  The latter ones are the most
dangerous for they in principle perturb the logic of the white noise
closure by introducing new independent equations and henceforth the
need for more renormalisation constants in the non diagonal sector of
the moments. Nevertheless one can argue a priori in the spirit of the
renormalisation group \cite{LeBellac}, only the nearest neighbours
interactions are relevant for scaling. Hence first order
corrections can be obtained allowing an $\epsilon$ dependence in the
renormalisation constant of the white noise closure and determining
the first order coefficient of their Taylor expansion. Moreover for
$\omega$ larger than two such a strategy is already able to take into
account the corrections due to the purely second neighbours
interactions.

\begin{figure}
  \begin{center}
  \epsfig{file=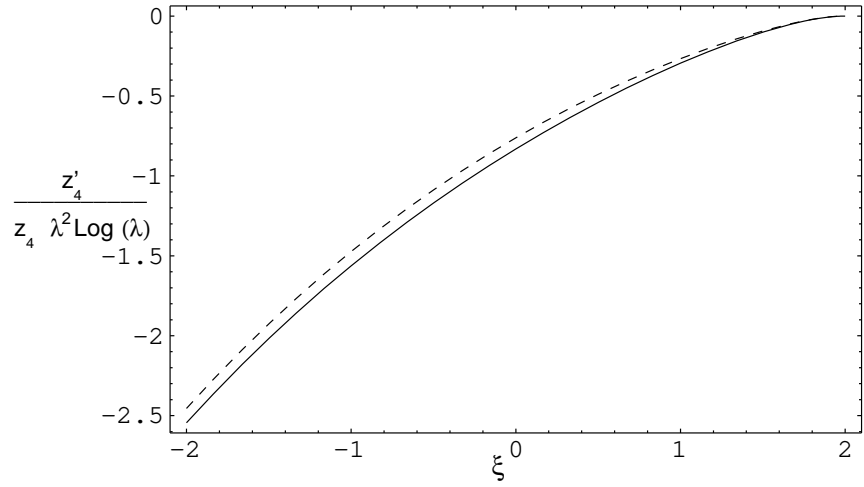}
   \vspace{0.5cm}
  \caption{the first order corrections to $H(4)$ with (dashed line) and 
    without inclusion of second neighbours interactions are plotted versus 
    the turbulence exponent. The inclusion of second neighbours couplings 
    increases the intensity of the anomaly}
  \label{fig:ouanomalfour}
 \end{center}
\end{figure}

Let us analyse in more detail the case of $C_{m,n}^{(4)}$. The white noise 
closure is perturbed by introducing an $\epsilon$ dependence in the 
renormalisation constants
\begin{eqnarray}
C_{m+n,m+n}^{(4)}&=&z(\epsilon)^{-n}C_{m,m}^{(4)}
\label{4perturbeddiag}\\
C_{m+n,m}^{(4)}&=&x(\epsilon)\,\lambda^{-H(2)}_{n-1}\,C_{m,m}^{(4)}.
\label{4perturbednondiag}
\end{eqnarray}
The marginal scaling in the non-diagonal sector in (\ref{4perturbednondiag}) 
is assumed to stay universal as it is for $C^{(2)}$ while the $\epsilon$ 
dependence is stored in the pre-factor.
The diagonal exponent is then determined up to first order as
\begin{equation}
H(4,\epsilon)=\frac{\log(z)}{\log(\lambda)}+\epsilon \lambda^2 \frac{z'}
{\lambda^2\,z \log(\lambda)}
\label{fourperturbed}
\end{equation}
where $z'$ is the derivative of $z$ at $\epsilon$ equal zero  
yielded by the perturbative solution of the system
\begin{equation}
  \left\{%
    \begin{array}{l}
\sum_{p,q}[\,I_{m,m;p,q}^{(4;0)}+\epsilon\,I_{m,m;p,q}^{(4;1)}]\,
C_{p,q}^{(4)}(\epsilon) =0\nonumber\\
\sum_{p,q}[\,I_{m,m-1;p,q}^{(4;0)}+\epsilon\,I_{m,m-1;p,q}^{(4;1)}\,
C_{p,q}^{(4)}(\epsilon)=0.
\label{4closedpert}
    \end{array}
  \right.
\end{equation}
The correction to $\rho_4$ due to time correlation increases 
the anomaly leading to a slower decay of the diagonal moment. For $z'$ is 
negative (see Fig.~\ref{fig:ouanomalfour}) the overall anomaly is
\begin{equation}
\rho_{4}(\epsilon)=(2-\xi)-\frac{\log(z)}{\log(\lambda)}+\left|\epsilon 
\lambda^2 \frac{z'}{\lambda^2\,z \log(\lambda)}\right|
\end{equation}
In the range of reliability of first order perturbation theory the effect 
is very small: for $\epsilon \lambda^2 \approx O(10^{-1})$ the prediction 
is a correction amounting to the three percent of the white noise exponent 
$H(4)$.
The perturbative scheme just proposed does not take into account the 
emergence of pure second neighbours interactions. In order to weight their 
relevance for the diagonal scaling and simultaneously to check the hypothesis 
of normal scaling for the marginal scaling in (\ref{4perturbednondiag})
one can relax the closure in order to encompass the equation
\begin{equation}
\sum_{p,q}[\,I_{m,m-2;p,q}^{(4;0)}+\epsilon\,I_{m,m-2;p,q}^{(4;1)}]\,
C_{p,q}^{(4)}(\epsilon)=0
\label{secondneighbours}
\end{equation}
which describe the independent, in the sense stated above, second neighbours
interaction. Consistency with the white noise theory imposes the latter 
equation to decouple when $\epsilon$ is set to zero. The requirement is
satisfied if the closure is chosen in the form
\begin{equation}
C_{m+n,m}^{(4)}=x(\epsilon) q(\epsilon)^{\frac{(n-1)(n-2)}{2}}\,
k_{(n-1)}{-H(2)}\,C_{m,m}^{(4)}
\end{equation}
the pre-factor $q(\epsilon)^{\frac{(n-1)(n-2)}{2}}$ enforces $q(0)$ to be 
a function of the white noise renormalisation constants. Were the white 
noise closure exact it would fix the value of $q(0)$ to one.

\begin{figure}
 \begin{center}
  \epsfig{file=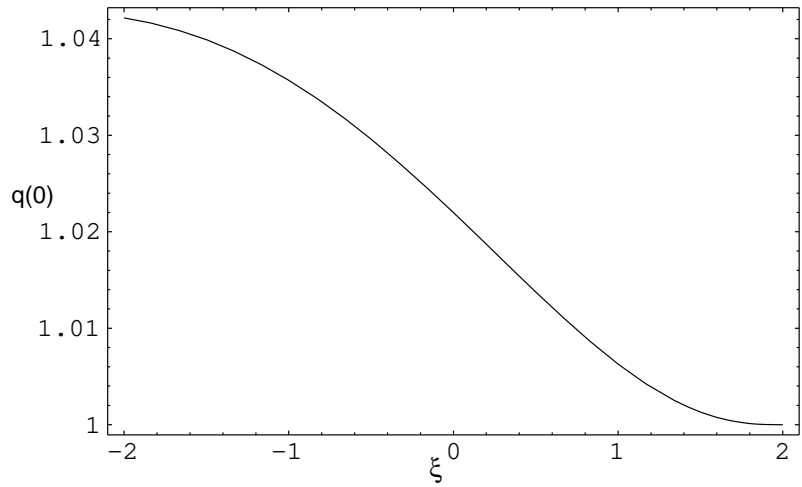}
   \vspace{0.5cm}
  \caption{The renormalisation constant $q(0)$ is plotted versus $\xi$.
     It remains close to one through all physical range proving self 
     consistent the conjecture of normal scaling for the non-diagonal 
     sector of the fourth moment. The result stresses that the 
     renormalisation of nearest neighbours interactions provide the 
     an accurate framework to extract the scaling exponent.}
  \label{fig:qzero}
 \end{center}
\end{figure}

In Fig. \ref{fig:qzero} the $q(0)$ is plotted versus $\xi$: through all the 
physical range it stays close to one with a maximal deviation on the
order of four percent for $\xi$ equal to minus two. 
Moreover as shown in Fig.~\ref{fig:ouanomalfour} 
the time correlation induced correction to $H(4,\epsilon)$ when 
(\ref{secondneighbours}) is included has the same qualitative behaviour 
and is quantitatively very close to the nearest-neighbours prediction.
The result is an a-posteriori check of the robustness of the closure 
approach. It confirms that first order corrections can be extracted 
within the logical scheme of the zero order one.
It follows that the equations specifying the zero modes of the inertial
operator acting on the sixth moment (see appendix~\ref{ap:sixthorder})
can be closed by assuming:
\begin{eqnarray}
C_{m+n,m+n,m+n}^{(6)}&=&z(\epsilon)^{-l}C_{m,m,m}^{(6)}\nonumber\\
C_{m+n,m+n,m}^{(6)}&=&x_{1}(\epsilon) k^{-H(4,\epsilon)}_{n-1}  
C_{m,m,m,m}^{(6)}\nonumber\\
C_{m+n+p,m+n,m}^{(6)}&=&x_{3}(\epsilon) k^{-H(2)}_{p-1} 
k^{- H(4,\epsilon)}_{n-1}  C_{m,m,m,m}^{(6)}\nonumber\\
C_{m+n,m,m,m}^{(6)}&=&x_{2}(\epsilon)k^{- H(2)}_{n-1} 
C_{m,m,m,m}^{(6)}.
\label{6scalingepsilon}
\end{eqnarray}
The exponent $H(4,\epsilon)$ is known perturbatively from 
(\ref{fourperturbed}) while $H(2)$ is universal.
With the same rationale (appendix~\ref{ap:sixthorder}) one can evaluate 
$H(8,\epsilon)$.

\begin{figure}[htbp]
 \begin{center}
  \epsfig{file=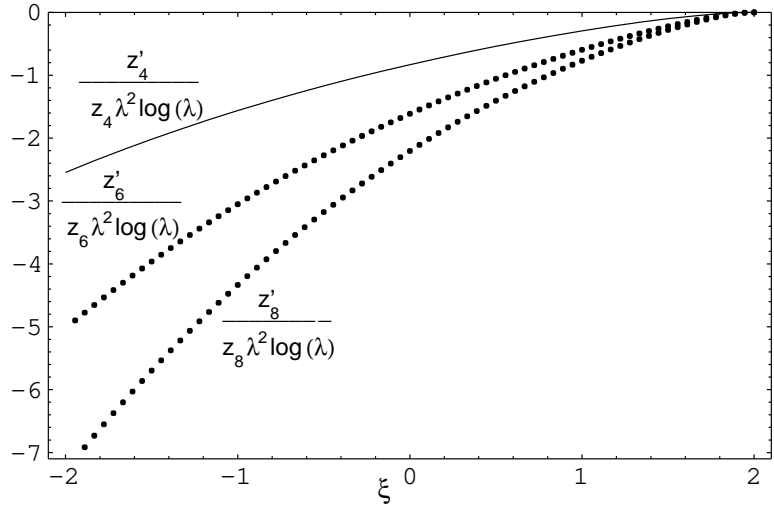}
   \vspace{0.5cm}
  \caption{The first order correction to $H(4)$ (continuous line), $H(6)$ and 
$H(8)$ versus the turbulence parameter $\xi$. In the last two cases 
the corresponding linear systems are solved numerically. 
In all cases the corrections are derived by perturbing the white noise closure 
``renormalisation'' constants. The effect of time correlation is seen to
add a negative correction to the scaling exponents highlighting an increase 
of intermittency.}
  \label{fig:ouanomal}
 \end{center}
\end{figure}

In Fig.~\ref{fig:ouanomal} the analytic predictions for the corrections 
to the scaling exponents are summarised. 
In all cases the corrections are negative i.e. they carry a positive 
contribution to $\rho_{2 \omega}$. 
The corrections increase with $\omega$, the rate of the growth being 
slightly slower than the 
$\Delta(2 \omega)\propto \omega(\omega-1)\,\Delta(4)/2 $ predicted in 
\cite{Chertkov} for time-correlation generalised PDE Kraichnan models.

Within the range of first order perturbation theory the overall effect 
of time correlation is seen  to enhance intermittency. 
An intuitive understanding of the phenomenon might be obtained by
interpreting the time correlation as a mechanism to increase the probability
of coherent fluctuations of the scalar field. The latter are rare events
felt in the tail of the probability density of the scalar field as extreme
deviations from the Gaussian behaviour of the typical events.

\section{Numerical experiments}
\label{s:numerics}
The resort to numerical experiments has a double motivation. On one
hand they can be used to test the predictions from the first order
perturbation theory.  On the other hand they provide a broader
scenario of the features of the model beyond the grasp of perturbative
approaches.  The first task is far from being easy because a
quantitative check of perturbation theory requires measurements of the
scaling exponents within an accuracy smaller than two percent.

The main feature of the inertial range is the conservation of the
scalar ``energy''.  From the analytical point of view this is seen in
the non-commutativity of the terms associated with the multiplicative
noise, $B^\gamma$ in (\ref{passcomplete}). This property rules out the
use of a simple Euler scheme, which can be applied in the case of
delta correlated noise. In the case of white noise advection the
multiplicative structure of the noise (\ref{passcomplete}) , which is
interpreted in the Stratonovich sense, can be mapped into the
corresponding It\^{o} equations.  The advantage is that the diagonal
non zero average part of the noise is explicitly turned into an
effective drift term \cite{Kloeden}. The non-diagonal terms in the
Taylor expansion of the scalar field $\Theta$ are of the order three
halves in $dt$, which are neglected in the Euler scheme.  This
procedure becomes meaningless for a time-correlated noise.  There
ordinary calculus holds and in the Taylor expansion of $\Theta$ both
diagonal and non-diagonal products of the noise are of the same order
in $dt$.

Moreover the algorithm to be used must tend smoothly to a 
white noise limit, so that the same relative error is preserved
independently on the value of $\epsilon$.

Following Burrage \& Burrage \cite{Burrage} a reliable way out from
the mentioned difficulties is to adopt the Trotter-Lie-Magnus formula
to integrate the equations of motion to first order.  For each time
increment $dt$ (\ref{passcomplete}) is solved in exponential form.
Fast matrix exponentiation algorithms are provided by the package
{\sc expokit} \cite{Sidje}. 

To generate the correlated noise, the exact method described by Miguel
\& Toral \cite{SanMiguel} is used. This method ensures that the noise is 
accurate down to the limit $\epsilon \rightarrow 0$.

\begin{figure}[htbp]
 \begin{center}
  \epsfig{file=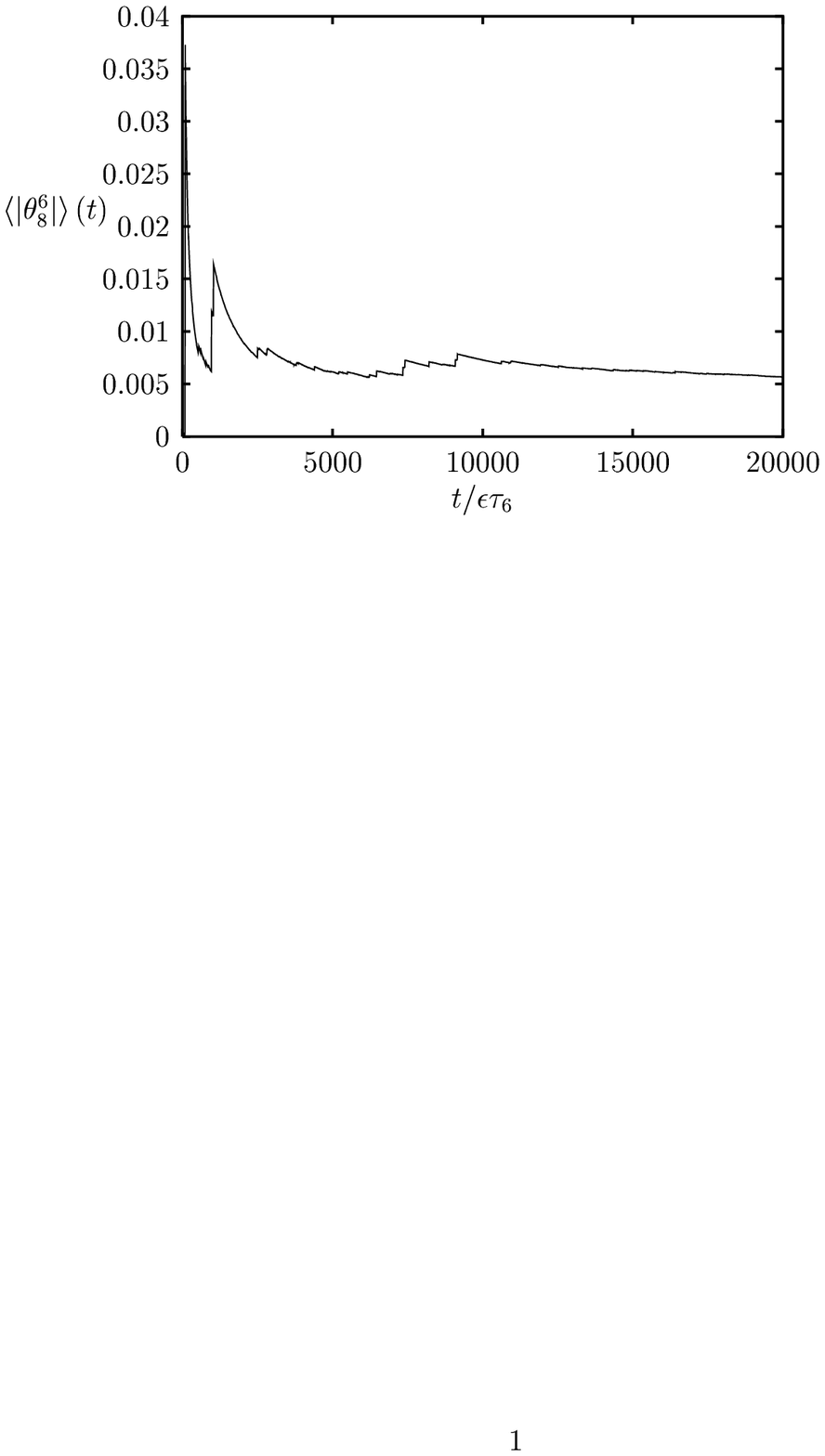}
  \caption{The convergence of the sixth order structure
  function for $\epsilon=1$. Shown is $\left<|\theta_m^6|\right>(t)$ for
  $m=8$. The fast upwards changes and slow downward relaxations reveal 
  the intermittent nature of the signal $\theta_{m=8}^6(t)$}
  \label{fig:contexample2}
 \end{center}
\end{figure}

The relevant time scale to measure the convergence of the solution is
the slowest time scale in the system, namely the eddy-turn-over time
of the first shell estimated as the maximum between $\epsilon\tau_1$
and $\tau_1$.  As shown in Fig.~\ref{fig:contexample2} more than
$N_\tau = 10000$ eddy-turn-over times are needed to achieve a
converged solution for the sixth order structure function. The time
step is set by the fastest time scale in the system, which is the one
of the largest shell $\epsilon \tau_M$. The number of iterations
needed to achieve convergence is then for $\epsilon$ less than one:
\begin{equation}
  \#(\mbox{iterations}) = \frac{N_\tau \tau_M}{\epsilon \tau_1} =
  \frac{N_\tau}{\epsilon} \lambda^{(M-1)(1-\frac{\xi}{2})}
\end{equation}
which shows that the number of iterations needed grows like $1/\epsilon$,
making it difficult to get close to the white noise limit using the same
algorithm.

The scaling of the diagonal moments of higher order has been extracted
by means of extended self-similarity \cite{ESS}, where the $p$-th order
structure function is plotted versus the second order one, which is
assumed to be normal. The scaling is found as the average slope of the
logarithmic derivatives in the inertial range.

We considered a system with twenty-five shells with wave numbers
increasing as power of $\lambda=2$, with viscosity $\kappa = 5 \times
10^{-9}$. This choice ensures that there are several shells in the
dissipative range. We focused on the results for $\xi$ equal two
thirds (Kolmogorov scaling).

\begin{figure}[tbp]
  \epsfig{file=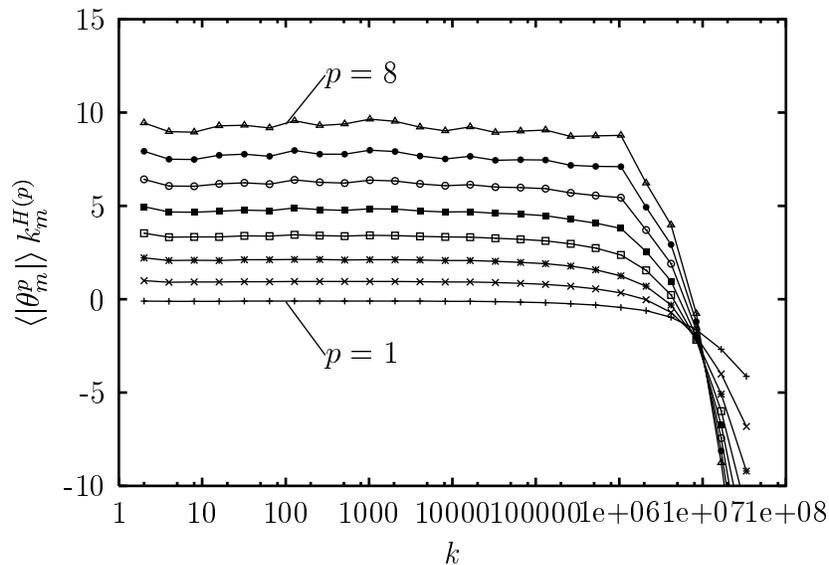}
  \caption{An example of the scaling of the structure functions for
    $\epsilon = 2.0$. The plot shows the structure functions
    ``normalised'' by the fitted scaling $k_m^{H(p)}$ to make the
      scaling regime appear as horizontal lines. The lower line is for
      $p=1$, and the upper line is $p=8$. Each line is offset to make
      it possible to distinguish the lines from each other.}
  \label{fig:structexample2}
\end{figure}

In Fig.~\ref{fig:structexample2} the ``normalised'' structure
functions $\left< | \theta_m^p | \right> k_m^{H(p)}$ are shown. The
quality of the scaling is demonstrated by the fact that the moments
show scaling over a wide range of scales.

\begin{figure}[tbp]
  \begin{center}
    \epsfig{file=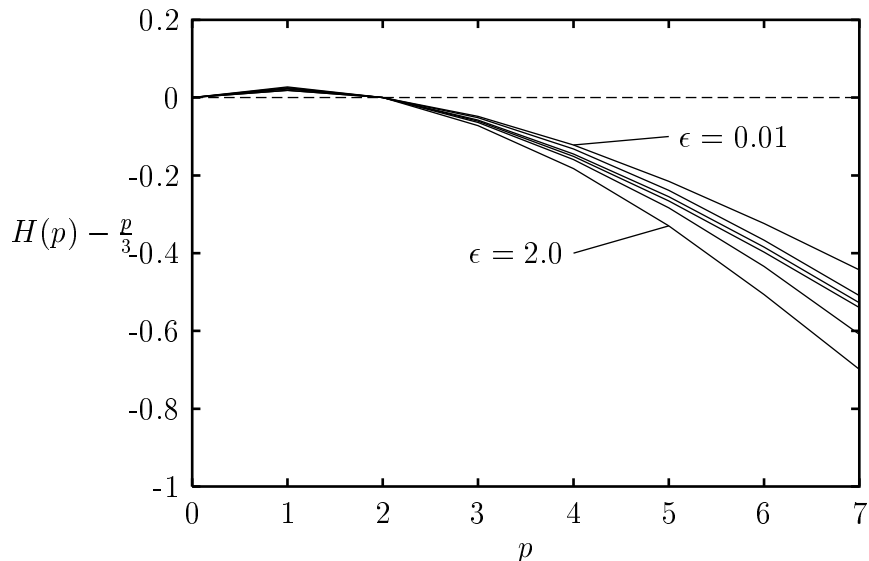}
    \caption{The anomalous part of the structure functions $H(p)-p/3$ 
      as a function of $p$ for $\epsilon = 0.01$ to $\epsilon=2$. Also
      shown are the points from the analytical calculation of the
      scaling for the white noise case. The lines correspond to (from the top):
      $\epsilon=0.01$, 0.02, 0.10, 0.25, 1.0 and 2.0. The dashed line
      corresponds to normal scaling.}
    \label{fig:compare}
  \end{center}
\end{figure}

A summary of the numerical experiments is given in
Fig.~\ref{fig:compare} where the scaling exponents are plotted
versus the order of the moments of the scalar field for different 
values of $\epsilon$. It is evident that the anomaly grows as the
time correlation increases.

When turning to the interpretation of the results in more detail, the
uncertainty in the extraction of the scaling has to be kept in mind.
For the sixth moment this uncertainty turned out to be on the order of
a four percent. The changes in the scaling between different values of
$\epsilon$ is also on the order of a few percent. This seems to
exclude a proper resolution in the numerics to compare the results
with the analytical predictions from the perturbation analysis.
However, the results for different values of $\epsilon$ can still be
compared with some confidence, as the relative uncertainty between the
different runs is much smaller than the absolute uncertainty. This
means that the slope of, i.e., the sixth order structure functions vs.
$\epsilon$ will be well resolved, while the absolute values can be
shifted up and down a few percent.

\begin{figure}[tbp]
  \begin{center}
    \epsfig{file=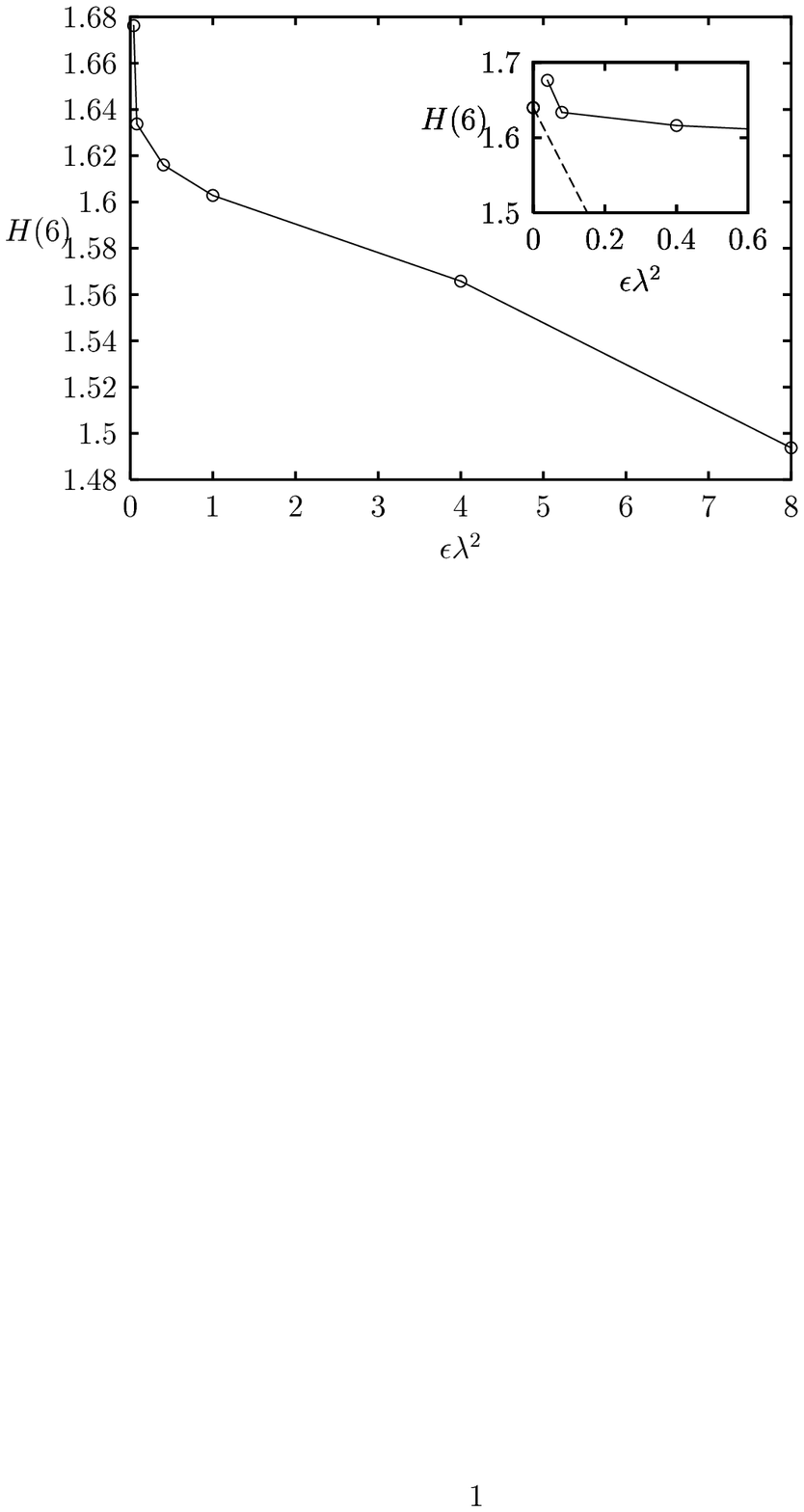}
    \caption{The scaling of the 6.th order structure function
      vs. $\epsilon \lambda^2$. The inset shows an enlargement of
      the perturbative range $\epsilon \lambda^2 << 1$ where the
      analytical prediction from first order perturbation theory
      can be compared with the numerical experiments.}
    \label{fig:epsilon}
  \end{center}
\end{figure}

In Fig.~\ref{fig:epsilon} the analytical prediction of the exponents
is compared with the result of the numerics. The theoretical points
are systematically below the numerical ones which is due to the
absolute uncertainty as explained above. The slope is the same for the
analytical calculation and the numerics, giving credibility to the
results of the perturbation analysis. It should be noted that the
effect of time correlation on the anomaly is quantitatively quite
small even in the non perturbative range when $\epsilon$ is equal to
one ($\epsilon \lambda^2$ equal four).

The global picture provided by the numerical experiments is that $H(2
\omega)$ is seen to be a non linear function of $\epsilon$ which,
after rapid initial decrease in the perturbative range, displays a
much slower rate of variation.  An interesting question is whether
there is a limiting value of the scaling of the structure function as
$\epsilon \gg 1$ or not. However the quality of the numerics does not
allow us to answer this question.

\section{Conclusion}
We have presented a shell model for the advection of a passive scalar
by a velocity field which is exponentially correlated in time. We
developed a systematic procedure to calculate the exponents of the
correlation of the diagonal moments (the structure functions). For the
delta correlated velocity we find good agreement between analytical
and numerical calculations up to the eight order.
We presented an analytical perturbative theory to compute the
correction to the scaling exponents due to the exponentially
correlated velocity field.

The occurrence of anomalies in the exponents of the diagonal moments
of the scalar and their non universality versus the intensity
$\epsilon\lambda^2$ of the time correlation, is related to the
presence of pure short range couplings in the corresponding inertial
operator which provide for non trivial scaling of the zero modes.  In
the absence of such short range couplings, as is the case for the
second moment, normal scaling would take place independently on the
value of $\epsilon \lambda^2$.

The behaviour of the anomalous exponents in the non-perturbative
regime was studied numerically. This was found to be a non linear
monotonic function of $\epsilon \lambda^2$, decreasing with a rate
much slower than in the perturbative regime. It is thus clear that the
addition of the time correlation to the advecting velocity field
enhances the anomalous scaling. The anomaly found in the present study
is still much smaller that what is found when the passive scalar is
driven by a turbulent velocity field driven by Navier-Stokes
turbulence or by a shell model for the velocity field \cite{Vulpio}.
This indicates that the non-Gaussian nature of the real turbulent
velocity field plays a significant r\^ole for the strong anomalous
scaling observed for real passive scalars.

\section{Acknowledgements}
The authors wish to thank A. Vulpiani for drawing their attention to the 
problem. 

Discussions with E. Aurell, A. Celani, P. Dimon and 
E. Henry are gratefully acknowledged. A particular thank to M. H. Jensen 
for his interest and encouragement during our work and to M. van Hecke
for many physically insightful comments.

PMG is supported by the TMR grant ERB4001GT962476 from the European 
Commission.

\appendix

\section{Stochastic Integration by Parts Formula}
\label{ap:one}

A heuristic proof of the stochastic integration by parts formula
is provided. For a rigorous treatment see \cite{Bass}, \cite{Nualart}.
Let $\zeta_t$ be a stochastic process whose realisations are defined as the 
solution of the It\^{o} SDE
\begin{eqnarray}
\begin{array}{lll}
\dot{x_t}=b(x_t,t)+\sigma(x_t,t)\eta_t\,, & \quad & x_t|_{t=0}=x_0
\end{array}
\label{a1sde1}
\end{eqnarray}
where $\eta_t$ is white noise.
Let $\zeta^{\epsilon}_t$ the stochastic process specified by 
\begin{eqnarray}
\begin{array}{lll}
\dot{x}_t=b(x_t,t)+\epsilon h(x_t,t)\sigma(x_t,t)+\sigma(x_t,t)\eta_t  
& \quad &  x_t|_{t=0}=x_0
\end{array}
\label{a1sde2}
\end{eqnarray}
For $\epsilon$ equal the two SDE's coalesce: (\ref{a1sde2}) can be derived 
from (\ref{a1sde1}) under the variation of the white noise
$\eta_t \rightarrow \eta_t+h(x_t,t)$. The integration by parts formula 
states that for any smooth functional $f$ the identity holds:
\begin{equation}
\left<\frac{d}{d\epsilon} f(\zeta^{\epsilon}_t)\right>_{\zeta^{\epsilon}_t}
|_{\epsilon=0}=\left<f(\zeta_t) \int_0^t ds\,h(\zeta_s,s) \right>_{\zeta_t}
\label{a1intbyparts}
\end{equation}
where $\left<...\right>_{\zeta_t}$ denotes the expectation values with respect 
to the measure induced by $\zeta_t$
In order to prove it let us observe that the transition probability density
for (\ref{a1sde2}) can be written formally as a path integral (It\^{o} 
discretisation):
\begin{eqnarray}
p_{\eta^{\epsilon}}(x,t|x_0,0)&=&\int_{x_0}^{x_t=x}{\cal D}x \,e^{-S_{\zeta}
(x,t|x_0,0)+\int_0^t\,dt' [\frac{\dot{x_{t'}}-b(x_{t'})}{\sigma(x_{t'},t')} 
\epsilon h(x_{t'},t')-\frac{\epsilon^2}{2} h^2(x_{t'},t')]}\nonumber\\
S_{\zeta}(x,t|x_0,0)&=&\int_{0}^{t} \,\,\frac{dt'}{2}
[\frac{\dot{x_{t'}}-b(x_{t'})}{\sigma(x_{t}',t')}]^2
\end{eqnarray}
If one introduces the functional
\begin{equation}
M(\zeta^{\epsilon}_t)=e^{-\int_{0}^{t}\,dt' [\frac{\dot{x_t'}-b(x_t',t')}
{\sigma(x_t',t')} \epsilon h(x_t',t')-\frac{\epsilon^2}{2}h^2(x_t',t')]}
\label{functional}
\end{equation}
one has by construction
\begin{equation}
\frac{d}{d\epsilon}\left<M(\zeta^{\epsilon}_t)f(\zeta^{\epsilon}_t)\right>
_{\zeta^{\epsilon}_t}=0
\end{equation}
To each realisation of the solutions of (\ref{a1sde2}) corresponds a mapping 
$\eta_t \rightarrow x_t=x(t,\eta_t,\epsilon)$. Hence the last equality 
can be rewritten as the white noise average:
\begin{equation}
\frac{d}{d\epsilon}\left<M(x(t,\eta_t,\epsilon))f(x(t,\eta_t,\epsilon)
)\right>_{\eta_t}=0
\end{equation}
which implies (\ref{a1intbyparts}) when $\epsilon$ is set to zero.
The derivative 
\begin{equation}
\frac{d}{d\epsilon} f(\zeta^{\epsilon}_t)|_{\epsilon=0}=
D \zeta_t\partial_{\zeta_t} f(\zeta_t)
\label{Frechet}
\end{equation}
is a Fr\'echet derivative. The dynamics of the stochastic process $D
\zeta_t$ is linear once the realisations $x_t$ of $\zeta_t$ are known:
\begin{eqnarray}
y_t &\equiv&D x_t \nonumber\\
\dot{y}_t&=& y_t \partial_{x_t}[b(x_t,t)+\sigma(x_t,t)\eta_t]+h(x_t,t)
\sigma(x_t,t)
\label{variation}
\end{eqnarray}
It is worth to note that for $b=0,\,\,\,\sigma=h=1$ the integration by parts 
formula (\ref{a1intbyparts}) reduces to
\begin{equation}
t \left<\partial_{w_t}f(w_t)\right>=\left<f(w_t)w_t \right>
\end{equation}
which is the Gaussian integration by parts formula (see e.g. \cite{Frisch}) 
applied to the Wiener process ${\cal N}(0,t)$. 

The generalisation to a multidimensional complex case proceeds 
straightforwardly by introducing $2N$ variational parameters 
$\{\epsilon_i,\epsilon_i^*\}_{i=1}^{2N}$ and applying the definitions
\begin{equation}
<\eta_{m}(t)\eta_{n}^{*}(s)>=2\,\delta_{mn}\delta(t-s) 
\end{equation}
for the white noise correlations.
 
\section{Stochastic integration by parts for the O-U process}
\label{ap:two}

As in the above appendix we limit ourselves to the real case the generalisation
to the complex case being straightforward.
Functional differentiation is formally derived from a Fr\'echet 
derivative with $h(x_t,t)=\delta(t-s)$ where $s$ is a parameter 
specifying the time when the white noise is perturbed. The variation 
is assumed to be non-anticipating (causal):
\begin{equation}
\lim_{s \uparrow t}\int_{0}^{t}\,ds'\delta(s-s')=0
\label{consistency}
\end{equation}
Let us consider the system of SDE's 
\begin{equation}
\dot{x}_m=b_m(x)+\sum_{n=1}^{2}\sigma_{m,n}(x)c_n(t)
\label{a2system}
\end{equation}
where $c$ is the coloured noise:
\begin{equation}
c_n(t)=\int_0^t\,ds'\, \frac{e^{-\frac{t-s'}
{\epsilon \tau_n}}}{\epsilon \sqrt{\tau_n}}\,\eta_n(s')
\label{coulored}
\end{equation}
Functional differentiation gives
\begin{equation}
\frac{d}{dt}(D^s_l x_m)=\sum_{k=1}^{N}D^s_l x_k
\,[\,\partial_k b_m(x)+\partial_k \sum_{n=1}^{N}
\sigma_{m,n}(x)\int_0^t\,ds' \frac{e^{-\frac{t-s'}
{\epsilon \tau_n}}}{\epsilon \sqrt{\tau_n}}\eta_n(s')]+\frac{e^{-\frac{t-s}
{\epsilon \tau_l}}}{\epsilon \sqrt{\tau_l}}\sigma_{m,l}(x)
\label{a2var}
\end{equation}
The functional derivative is fully specified when it is known its form at 
the time $s$ when the variation of the white noise occurs. The latter is
determined by the causality requirement 
\begin{equation}
\frac{d}{dt}(D^s_l c_n(t))=[\,\partial_t \frac{e^{-\frac{t-s}
{\epsilon \tau_n}}}{\epsilon \sqrt{\tau_n}} +\frac{1}
{\epsilon \sqrt{\tau_n}}\delta(t-s)\,]\,
\delta_{n,l}
\label{varnoise}
\end{equation}
which implies the variation of the coloured noise to be nonzero {\em only} 
immediately after the instantaneous kick
\begin{equation} 
D^s_l c_n(t)=\frac{e^{-\frac{t-s}
{\epsilon \tau_n}}}{\epsilon \sqrt{\tau_n}}\delta_{n,l}\,\,\, \quad \forall t 
\geq s
\label{coloursol}
\end{equation}
By differentiating (\ref{a2var}) one finds
\begin{equation}
\frac{d^2}{dt^2}(D^s_l x_m)=\frac{e^{-\frac{t-s}
{\epsilon \tau_l}}}{\epsilon \sqrt{\tau_l}}\,\sigma_{m,l}(x)
\,\delta(t-s)+\mbox {smooth terms}
\label{a2var2derivative}
\end{equation}
>From the last equation it stems that for $t=s$ 
\begin{equation}
\frac{d}{dt}(D^s_l x_m)|_{t=s}=\frac{1}{\epsilon \sqrt{\tau_n}}
\sigma_{m,l}(x)
\end{equation}
Consistency with (\ref{a2var}) then requires that the variation of the 
$x$'s associated with a non anticipating variation of the white noise 
at time $s$ fulfils the initial condition:
\begin{equation}
D^s_l x_m(s)|_{t=s}=0
\end{equation}
The integration by parts formula (\ref{a1intbyparts}) for a smooth
functional $O(x)$ 
\begin{equation}
\left<O(x_t)c_n(t)\right>=
\int_0^t ds'\, \frac{e^{-\frac{t-s'}{\epsilon \tau_n}}}
{\epsilon \sqrt{\tau_n}}\sum_{l=1}^{N}\left<D^{s'}x_l
\partial_{x_l}\,O(x_t)\right>
\label{a2intbyparts}
\end{equation}
The variation is the solution of the linear problem (\ref{a2var}) of which 
we define $R$ to be the fundamental solution. It follows
\begin{equation}
\left<O(x_t)c_n(t)\right>=
\int_0^t ds\, \frac{e^{-\frac{t-s}{\epsilon \tau_n}}}{\epsilon}
\int_s^t ds'\,\frac{e^{-\frac{s'-s}{\epsilon \tau_n}}}{\epsilon}
\,\sum_{l=1}^{N}\sum_{m=1}^{N}
\left<[\partial_{x_l}\,O(x_t)]\,R_{l,m}(t,s')\sigma_{m,n}
(x_s')\right>
\end{equation}
Finally inverting the order of integration one obtains
\begin{equation}
\left<O(x_t)c_n(t)\right>=
\int_0^t ds'\, \frac{e^{-\frac{t-s'}{\epsilon \tau_n}}-
e^{-\frac{t+s'}{\epsilon \tau_n}}}{2 \epsilon}
\,\sum_{l=1}^{N}\sum_{m=1}^{N}
\left<[\partial_{x_l}\,O(x_t)]\,R_{l,m}(t,s')\sigma_{m,n}
(x_s')\right>
\label{result}
\end{equation}
This proves the real version of formula (\ref{typical}).

\section{The Fourth order correlation to first order}
\label{ap:fourthorder}
The inertial operator acting on the fourth moment $C_{m,n}^{(4)}(t)$ 
is in the large time limit 
\begin{eqnarray}
&&\lefteqn{RHS=I_{m,n;p,q}^{(4,0)}C^{(4)}_{p,q}-2k_{m+1}^2 d_{m}\int_{0}^{t}ds
\,
e^{-\frac{t-s}{\epsilon \tau_m}}\frac{d}{ds}\Re{\cal F}_{m,n}^{(4)}(t,s)+}
\nonumber\\
&&+k_{m}^2 d_{m-1}\int_{0}^{t}ds\,e^{-\frac{t-s}{\epsilon \tau_{m-1}}}
\frac{d}{ds}\Re{\cal F}_{m-1,n}^{(4)}(t,s)+
\nonumber\\
&&-k_{n+1}^2 d_{n}\int_{0}^{t}ds\,e^{-\frac{t-s}{\epsilon \tau_{n}}} 
\frac{d}{ds}\Re{\cal F}_{n,m}^{(4)}(t,s) \nonumber\\
&&+k_{n}^2 d_{n-1}\int_{0}^{t}ds\,e^{-\frac{t-s}{\epsilon \tau_{n-1}}} 
\frac{d}{ds}\Re{\cal F}_{n-1,m}^{(4)}(t,s)
\label{4pointbyparts}
\end{eqnarray}
The bi-dimensional matrix $I_{m,n;p,q}^{(4,0)}$ is the white noise 
linear inertial operator. The corrections to the white noise theory are 
generated by the time derivative at equal times of the integrand function 
$\Re{\cal F}_{n,m}^{(4)}$
\begin{eqnarray}
&&{\cal F}_{m,n}^{(4)}(t,s)\dot{=} 
    \left< \Theta_{N+m}(t) \Theta_{N+n}(t)\Theta_{n}(t)
    R_{N+m+1,N+m}(t, s)\Theta_{m+1}(s)]\right>+\nonumber\\
&&  -\left< \Theta_{N+m}(t) \Theta_{N+n}(t)\Theta_{n}(t)
    R_{N+m+1,N+m+1}(t, s)\Theta_{m}(s)]\right> +\nonumber\\
&&  + \left< \Theta_{N+m+1}(t) \Theta_{N+n}(t)\Theta_{n}(t)
    R_{N+m,N+m}(t, s)\Theta_{m+1}(s)]\right> +\nonumber\\
&&  -\left< \Theta_{N+m+1}(t) \Theta_{N+n}(t)\Theta_{n}(t)
    R_{N+m,N+m+1}(t, s)\Theta_{m}(s)]\right> +\nonumber\\
&&  + \left< \Theta_{N+m}(t) \Theta_{N+m+1}(t)\Theta_{N+n}(t)
    R_{n,N+m}(t, s)\Theta_{m+1}(s)]\right> +\nonumber\\
&&  - \left< \Theta_{N+m}(t) \Theta_{N+m+1}(t)\Theta_{N+n}(t)
    R_{n,N+m+1}(t, s)\Theta_{m}(s)]\right>  +\nonumber\\
&&  + \left< \Theta_{N+m}(t) \Theta_{N+m+1}(t)\Theta_{n}(t)
    R_{N+n,N+m}(t, s)\Theta_{m+1}(s)]\right> +\nonumber\\
&&  - \left< \Theta_{N+m}(t) \Theta_{N+m+1}(t)\Theta_{n}(t)
    R_{N+n,N+m+1}(t, s)\Theta_{m}(s)]\right> 
 \label{4tobeint}
\end{eqnarray}

After a double integration by parts neglecting viscous contributions one gets 
into
\begin{eqnarray}
&&\sum_{p,q} \frac{(I_{m,n;p,q}^{(4;0)}+\epsilon\,I_{m,n;p,q}^{(4;1)})
}{2}C^{(4)}_{q,p}= \nonumber\\
&&=  
   (  -{\frac{{{\lambda }^2}}{{{\tau }_{-1 + m}}}}  - 
       {\frac{{{\lambda }^2}}{{{\tau }_m}}} - 
       {\frac{{{\lambda }^2}}{{{\tau }_n}}}  - 
       {\frac{{{\lambda }^2}}{{{\tau }_{-1 + n}}}} +   
       {\frac{7\,\epsilon \,{{\lambda }^4}}{{{\tau }_{-1 + m}}}} + 
       {\frac{7\,\epsilon \,{{\lambda }^4}}{{{\tau }_m}}} + 
       {\frac{7\,\epsilon \,{{\lambda }^4}}{{{\tau }_{-1 + n}}}} + 
       {\frac{7\,\epsilon \,{{\lambda }^4}}{{{\tau }_n}}} 
   )C^{(4)}_{m,n}   + \nonumber\\
&&+(   {\frac{{{\lambda }^2}}{{{\tau }_m}}}- 
       {\frac{\epsilon \,{{\lambda }^4}}{{{\tau }_{1 + m}}}}  - 
       {\frac{2\,\epsilon \,{{\lambda }^4}}{{{\tau }_{-1 + n}}}} - 
       {\frac{2\,\epsilon \,{{\lambda }^4}}{{{\tau }_n}}} - 
       {\frac{7\,\epsilon \,{{\lambda }^4}}{{{\tau }_m}}} 
   )C^{(4)}_{m+1,n}\,+\nonumber\\
&&+(   {\frac{{{\lambda }^2}}{{{\tau }_n}}} - 
       {\frac{\epsilon \,{{\lambda }^4}}{{{\tau }_{1 + n}}}} -
       {\frac{2\,\epsilon \,{{\lambda }^4}}{{{\tau }_{-1 + m}}}} - 
       {\frac{2\,\epsilon \,{{\lambda }^4}}{{{\tau }_m}}} - 
       {\frac{7\,\epsilon \,{{\lambda }^4}}{{{\tau }_n}}}
   ) C^{(4)}_{m,n+1}+\nonumber\\
&&+(   {\frac{{{\lambda }^2}}{{{\tau }_{-1 + m}}}} -
       {\frac{\epsilon \,{{\lambda }^4}}{{{\tau }_{-2 + m}}}}  - 
       {\frac{2\,\epsilon \,{{\lambda }^4}}{{{\tau }_{-1 + n}}}} - 
       {\frac{2\,\epsilon \,{{\lambda }^4}}{{{\tau }_n}}} - 
       {\frac{7\,\epsilon \,{{\lambda }^4}}{{{\tau }_{-1 + m}}}} 
   )C^{(4)}_{m-1,n}\, + \nonumber\\
&&+(   {\frac{{{\lambda }^2}}{{{\tau }_{-1 + n}}}} - 
       {\frac{\epsilon \,{{\lambda }^4}}{{{\tau }_{-2 + n}}}} - 
       {\frac{2\,\epsilon \,{{\lambda }^4}}{{{\tau }_{-1 + m}}}} - 
       {\frac{2\,\epsilon \,{{\lambda }^4}}{{{\tau }_m}}}  - 
       {\frac{7\,\epsilon \,{{\lambda }^4}}{{{\tau }_{-1 + n}}}}
   )\, C^{(4)}_{m,n-1} + \nonumber\\
&&+(   {\frac{2\,\epsilon \,{{\lambda }^4}}{{{\tau }_m}}} + 
       {\frac{2\,\epsilon \,{{\lambda }^4}}{{{\tau }_n}}} 
   )\,C^{(4)}_{m+1,n+1}
  +(   {\frac{2\,\epsilon \,{{\lambda }^4}}{{{\tau }_{-1 + m}}}} + 
       {\frac{2\,\epsilon \,{{\lambda }^4}}{{{\tau }_{-1 + n}}}}
   )\,C^{(4)}_{m-1,n-1}  + \nonumber\\
&&+(   {\frac{2\,\epsilon \,{{\lambda }^4}}{{{\tau }_m}}} + 
       {\frac{2\,\epsilon \,{{\lambda }^4}}{{{\tau }_{-1 + n}}}}
   )\,C^{(4)}_{m+1,n-1}\, + 
   (   {\frac{2\,\epsilon \,{{\lambda }^4}}{{{\tau }_{-1 + m}}}} + 
       {\frac{2\,\epsilon \,{{\lambda }^4}}{{{\tau }_n}}} 
   ) \,C^{(4)}_{m-1,n+1}+ \nonumber\\
&&+    {\frac{\epsilon \,{{\lambda }^4}}{{{\tau }_{-2 + m}}}}
\,C^{(4)}_{m-2,n} + 
       {\frac{\epsilon \,{{\lambda }^4}}{{{\tau }_{1 + m}}}}\,
C^{(4)}_{m+2,n} + 
       {\frac{\epsilon \,{{\lambda }^4}}{{{\tau }_{-2 + n}}}}\,
C^{(4)}_{m,n-2} +    
       {\frac{\epsilon \,{{\lambda }^4}}{{{\tau }_{1 + n}}}}\,C^{(4)}_{m,n+2} 
+\nonumber\\ 
&&+\delta_{n,m}[  
  (    {\frac{2\,{{\lambda }^2}}{{{\tau }_m}}} - 
       {\frac{2\,\epsilon \,{{\lambda }^4}}{{{\tau }_{1 + m}}}} -
       {\frac{4\,\epsilon \,{{\lambda }^4}}{{{\tau }_{-1 + m}}}} - 
       {\frac{34\,\epsilon \,{{\lambda }^4}}{{{\tau }_m}}}
  )\,C^{(4)}_{m+1,m}   + \nonumber\\
&&+(   {\frac{2\,{{\lambda }^2}}{{{\tau }_{-1 + m}}}} -
       {\frac{2\,\epsilon \,{{\lambda }^4}}{{{\tau }_{-2 + m}}}}  - 
       {\frac{4\,\epsilon \,{{\lambda }^4}}{{{\tau }_m}}}  - 
       {\frac{34\,\epsilon \,{{\lambda }^4}}{{{\tau }_{-1 + m}}}}
  )\, C^{(4)}_{m,m-1}+ \nonumber\\
&&+(     {\frac{4\,\epsilon \,{{\lambda }^4}}{{{\tau }_{-1 + m}}}} + 
       {\frac{4\,\epsilon \,{{\lambda }^4}}{{{\tau }_m}}} 
  )\,C^{(4)}_{m,m}+ 
 (     {\frac{4\,\epsilon \,{{\lambda }^4}}{{{\tau }_{-1 + m}}}} + 
       {\frac{4\,\epsilon \,{{\lambda }^4}}{{{\tau }_m}}} 
  )\,C^{(4)}_{m+1,m-1}\, + \nonumber\\
&&+    {\frac{4\,\epsilon \,{{\lambda }^4}}{{{\tau }_m}}}
   \,C^{(4)}_{m+1,m+1} +
       {\frac{4\,\epsilon \,{{\lambda }^4}}{{{\tau }_{-1 + m}}}}
   \,C^{(4)}_{m-1,m-1} + 
       {\frac{2\,\epsilon \,{{\lambda }^4}}{{{\tau }_{1 + m}}}}
   \,C^{(4)}_{m+2,m}+
       {\frac{2\,\epsilon \,{{\lambda }^4}}{{{\tau }_{-2 + m}}}}
   \,C^{(4)}_{m,m-2}]
+\nonumber\\
&&+\delta_{n,m+1}
[  ( - {\frac{2\,{{\lambda }^2}}{{{\tau }_m}}}+
       {\frac{3\,\epsilon \,{{\lambda }^4}}{{{\tau }_{-1 + m}}}} + 
       {\frac{3\,\epsilon \,{{\lambda }^4}}{{{\tau }_{1 + m}}}} + 
       {\frac{34\,\epsilon \,{{\lambda }^4}}{{{\tau }_m}}}
  )\,C^{(4)}_{m+1,m}-   
       {\frac{4\,\epsilon \,{{\lambda }^4}}{{{\tau }_m}}}
   \,C^{(4)}_{m,m} +\nonumber\\
&&+(   {\frac{3\,\epsilon \,{{\lambda }^4}}{{{\tau }_{-1 + m}}}} + 
       {\frac{\epsilon \,{{\lambda }^4}}{{{\tau }_m}}} 
  )\,C^{(4)}_{m,m-1}\,-    
  (    {\frac{3\,\epsilon \,{{\lambda }^4}}{{{\tau }_{-1 + m}}}}+    
       {\frac{\epsilon \,{{\lambda }^4}}{{{\tau }_m}}}
  )\,C^{(4)}_{m+1,m-1}-   
       {\frac{4\,\epsilon \,{{\lambda }^4}}{{{\tau }_m}}}
   \,C^{(4)}_{m+1,m+1} +\nonumber\\
&&+  ( {\frac{\epsilon \,{{\lambda }^4}}{{{\tau }_m}}} + 
       {\frac{3\,\epsilon \,{{\lambda }^4}}{{{\tau }_{1 + m}}}} 
  )\,C^{(4)}_{m+2,m+1}\, -
  (    {\frac{\epsilon \,{{\lambda }^4}}{{{\tau }_m}}}+ 
       {\frac{3\,\epsilon \,{{\lambda }^4}}{{{\tau }_{1 + m}}}}
  )\,C^{(4)}_{m+2,m}]+
\nonumber\\
&&+\delta_{n,m-1}[
  (  - {\frac{2\,{{\lambda }^2}}{{{\tau }_{-1 + m}}}}+ 
       {\frac{3\,\epsilon \,{{\lambda }^4}}{{{\tau }_{-2 + m}}}} + 
       {\frac{3\,\epsilon \,{{\lambda }^4}}{{{\tau }_m}}} + 
       {\frac{34\,\epsilon \,{{\lambda }^4}}{{{\tau }_{-1 + m}}}}
  )\,C^{(4)}_{m,m-1} - 
       {\frac{4\,\epsilon \,{{\lambda }^4}}{{{\tau }_{-1 + m}}}}
   \,C^{(4)}_{m,m}+  \nonumber\\
&&+(   {\frac{\epsilon \,{{\lambda }^4}}{{{\tau }_{-1 + m}}}} + 
       {\frac{3\,\epsilon \,{{\lambda }^4}}{{{\tau }_m}}} 
  )\,C^{(4)}_{m+1,m} - 
 (    {\frac{\epsilon \,{{\lambda }^4}}{{{\tau }_{-1 + m}}}} + 
       {\frac{3\,\epsilon \,{{\lambda }^4}}{{{\tau }_m}}} 
  )\,C^{(4)}_{m+1,m-1}-   
       {\frac{4\,\epsilon \,{{\lambda }^4}}{{{\tau }_{-1 + m}}}}
   \,C^{(4)}_{m-1,m-1} +\nonumber\\
&&+(   {\frac{3\,\epsilon \,{{\lambda }^4}}{{{\tau }_{-2 + m}}}} + 
       {\frac{\epsilon \,{{\lambda }^4}}{{{\tau }_{-1 + m}}}}
  )\,C^{(4)}_{m-1,m-2}\,-  
  (    {\frac{3\,\epsilon \,{{\lambda }^4}}{{{\tau }_{-2 + m}}}} + 
       {\frac{\epsilon \,{{\lambda }^4}}{{{\tau }_{-1 + m}}}}
  )\,C^{(4)}_{m,m-2}]+
\nonumber\\
&&+\delta_{n,m+2}[
 (    -{\frac{3\,\epsilon \,{{\lambda }^4}}{{{\tau }_m}}} - 
       {\frac{\epsilon \,{{\lambda }^4}}{{{\tau }_{1 + m}}}} 
 ) C^{(4)}_{m+1,m}\,
 (    -{\frac{\epsilon \,{{\lambda }^4}}{{{\tau }_m}}} - 
       {\frac{3\,\epsilon \,{{\lambda }^4}}{{{\tau }_{1 + m}}}} 
  )C^{(4)}_{m+2,m+1}  +\nonumber\\
&&+( {\frac{\epsilon \,{{\lambda }^4}}{{{\tau }_m}}} + 
     {\frac{\epsilon \,{{\lambda }^4}}{{{\tau }_{1 + m}}}} 
   )C^{(4)}_{m+2,m}\,]+
\nonumber\\
&&+\delta_{n,m-2}[
   ( -{\frac{\epsilon \,{{\lambda }^4}}{{{\tau }_{-2 + m}}}} - 
      {\frac{3\,\epsilon \,{{\lambda }^4}}{{{\tau }_{-1 + m}}}}
   )C^{(4)}_{m,m-1}\,  + 
   ( -{\frac{3\,\epsilon \,{{\lambda }^4}}{{{\tau }_{-2 + m}}}} - 
     {\frac{\epsilon \,{{\lambda }^4}}{{{\tau }_{-1 + m}}}}
   )C^{(4)}_{m-1,m-2} +\nonumber\\
&&+( {\frac{\epsilon \,{{\lambda }^4}}{{{\tau }_{-2 + m}}}} + 
     {\frac{\epsilon \,{{\lambda }^4}}{{{\tau }_{-1 + m}}}}
   ) C^{(4)}_{m,m-2}\,]
\label{fourthorder}
\end{eqnarray}
The diagonal scaling exponent is derived up to first order is $\epsilon$ 
resorting to linear perturbation theory. If pure second neighbours interactions
are taken into account the constant $q(0)$ is specified by
\begin{equation}
 q(0)=\frac{1 - \left( 1 + {{\lambda }^{-2\,H(2)}} + 
        {{\lambda }^{-H(2)}} \right) \,z(0)}{z(0) + 
     z(0)^2\,\lambda^{-3\,H(2)}}
\end{equation}
The result is approximately equal to one for all $\xi$ ranging between $[0,2]$.
The first order correction $z'(0)$ is extracted from the solution of the 
linear system:
\begin{eqnarray}
&&  ( 4\,{{\lambda }^{2 + H(2)}} + 
     4\,{{\lambda }^2}\,z(0) ) \,x'(0) + 
  4\,{{\lambda }^2}\,x(0)\,z'(0)=\nonumber\\
&&=    -4\,{{\lambda }^{4 + 2\,H(2)}}\,x(0) + 
  4\,{{\lambda }^{4 - H(2)}}\,x(0)\,z(0) + 
  4\,{{\lambda }^{2 + 2\,( 1 - H(2) ) }}\,x(0)\,
   {{z(0)}^2}+
\nonumber\\
&& + 2\,{{\lambda }^4}\,
   ( 9 - 4\,x(0) + ( 4 - 22\,x(0) ) \,z(0)
      )  + 2\,{{\lambda }^{4 + H(2)}}\,
      ( 4 + ( 9 - 24\,x(0) )  - 
        4\,x(0)\,{{z(0)}} )
\nonumber
\end{eqnarray}
\begin{eqnarray}
&&[ {{\lambda }^2}\,
   ( -3 - {{\lambda }^{-H(2)}} - 
     {{\lambda }^{H(2)}} ) \,z(0) + {{\lambda }^{2 - 2\,H(2)}}\,{{z(0)}^2}
] \,x'(0) +   \nonumber\\
&& + 
[ {{\lambda }^2} + ( -3 - {{\lambda }^{-H(2)}} - 
     {{\lambda }^{H(2)}} ) \,x(0){{\lambda }^2}   + 
2\,{{\lambda }^{2 - 2\,H(2)}}\,x(0)\,z(0) 
] \,z'(0)=\nonumber\\
&& =
-13\,{{\lambda }^4} - 3\,{{\lambda }^{4 + H(2)}} + 
  3\,{{\lambda }^{4 + H(2)}}\,
   ( 2 + {{\lambda }^{-H(2)}} ) \,x(0) - 
  13\,{{\lambda }^4}\,z(0) - 3\,{{\lambda }^{4 - H(2)}}\,z(0)+
\nonumber\\
&& + {{\lambda }^4}\,( 42 - 2\lambda^{-2\,H(2)} + 
     7\lambda^{-H(2)} + 9\,{{\lambda }^{H(2)}} + 
                  q(0) ) \,x(0)\,z(0)+
\nonumber\\
&& + {{\lambda }^4}\,x(0)\,{{z(0)}^2}[ 1 - 10\lambda^{-2\,H(2)} + 
      (-1 + 2\,q(0))\lambda^{-3\,H(2)} + 
             (3 + 2\,q(0))\lambda^{-H(2)} ]+
\nonumber\\
&&  + 
     {{\lambda }^{4 - 4\,H(2)}}\,q(0)\,x(0)\,{{z(0)}^3}
\nonumber
\end{eqnarray}
\begin{eqnarray}
&& {{\lambda }^{2 - H(2)}}\,x(0)\,[1  - 
     2\,z(0)\,( 1 - q(0)  + 3\,{{\lambda }^{- 2\,H(2)}}\,q(0)\,z(0)+ 
        {{\lambda }^{-2\,H(2)}}+ {{\lambda }^{-H(2)}})
      ] \,z'(0)  +
\nonumber\\
&& + {{\lambda }^{2 - H(2)}}\,z(0)\,[ 1 - 
    z(0)\,( 1 + {{\lambda }^{-2\,H(2)}} + 
        {{\lambda }^{-H(2)}} - q(0)  + 
     {{\lambda }^{- 2\,H(2)}}\,q(0)\,z(0)\,)
     ]  \,x'(0)+
\nonumber\\
&& +  {{\lambda }^{2 - H(2)}}\,{{z(0)}^2\, 
     [ x(0)} + {{\lambda }^{- 2\,H(2)}}\,x(0)\,z(0)
     ]  \,q'(0) =
\nonumber\\
&&  =  2\,{{\lambda }^{4 + H(2)}}\,
   ( -1 + {{\lambda }^{-3\,H(2)}} - 
     4\lambda^{-2\,H(2)} - 2\lambda^{-H(2)} ) \,x(0)\,z(0) +
\nonumber\\
&&+ 
  {{\lambda }^{4 - H(2)}}\,{{z(0)}^2} + 
  6\,{{\lambda }^{2 + 2\,( 1 - H(2) ) }}\,x(0)\,
   {{z(0)}^2}- 2\,{{\lambda }^{4 - 4\,H(2)}}\,q(0)\,x(0)\,
   {{z(0)}^2}+
\nonumber\\
&&  - {{\lambda }^4}\,( 1 + q(0) ) \,x(0)\,
   {{z(0)}^2} - {{\lambda }^{4 - 3\,H(2)}}\,
   ( -7 + 2\,q(0) ) \,x(0)\,{{z(0)}^2}+
\nonumber\\
&& + 
  {{\lambda }^{4 - H(2)}}\,
   ( 2 - 7\,q(0) + {{q(0)}^3} ) \,x(0)\,{{z(0)}^2} + 
  2\,{{\lambda }^{4 - H(2)}}\,
   ( 1 + {{\lambda }^{-2\,H(2)}} ) \,x(0)\,{{z(0)}^3}+
\nonumber\\
&&- {{\lambda }^{4 - H(2)}}\,
   ( 2 + {{\lambda }^{-4\,H(2)}} + 
     7\lambda^{3\,H(2)}+{{\lambda }^4} + 2\,{{\lambda }^4}\,
   ( 1 + {{\lambda }^{-H(2)}} ) \,z(0)+
\nonumber\\
&& + 2\lambda^{H(2)} ) \,q(0)\,x(0)\,
   {{z(0)}^3} + 2\,{{\lambda }^{4 - 2\,H(2)}}\,
   ( 1 + {{\lambda }^{-3\,H(2)}} ) \,{{q(0)}^3}\,
   x(0)\,{{z(0)}^3}+
\nonumber\\
&& + {{\lambda }^{4 - 6\,H(2)}}\,{{q(0)}^3}\,
     x(0)\,{{z(0)}^4}
\end{eqnarray}

\section{The inertial operator for the sixth moment of the correlation up 
         to first order} 
\label{ap:sixthorder}

Under the hypothesis that pure second neighbours interactions do not require
new equations to specify the diagonal scaling for small values of $\epsilon$,
there are only four equations describing how global coupling are 
``renormalised'' by relevant pure short range interactions.  
Given the $m$-th shell one has 
\begin{eqnarray}
&&0=\sum_{p,q,r}[I_{m,m,m;p,q,r}^{(6;0)}+\epsilon I_{m,m,m;p,q,r}^{(6;1)}]\,
C_{p,q,r}^{(6);1}=\nonumber\\
&&=  
 - ( {\frac{3\,{{\lambda }^2}}{{{\tau }_{-1 + m}}}} + 
     {\frac{3\,{{\lambda }^2}}{{{\tau }_m}}} - 
     {\frac{45\,\epsilon \,{{\lambda }^4}}{{{\tau }_{-1 + m}}}} - 
     {\frac{45\,\epsilon \,{{\lambda }^4}}{{{\tau }_m}}} 
  )\,C^{(6)}_{m,m,m}  +\nonumber\\
&&   + 
  (  {\frac{9\,{{\lambda }^2}}{{{\tau }_m}}} - 
     {\frac{9\,\epsilon \,{{\lambda }^4}}{{{\tau }_{1 + m}}}} -
     {\frac{36\,\epsilon \,{{\lambda }^4}}{{{\tau }_{-1 + m}}}} - 
     {\frac{207\,\epsilon \,{{\lambda }^4}}{{{\tau }_m}}}
  )\,C^{(6)}_{m,m,1 + m} +\nonumber\\
&& + 
  (  {\frac{9\,{{\lambda }^2}}{{{\tau }_{-1 + m}}}}-
     {\frac{9\,\epsilon \,{{\lambda }^4}}{{{\tau }_{-2 + m}}}} - 
     {\frac{36\,\epsilon \,{{\lambda }^4}}{{{\tau }_m}}}  - 
     {\frac{207\,\epsilon \,{{\lambda }^4}}{{{\tau }_{-1 + m}}}}
  )\,C^{(6)}_{m,m,-1 + m}+\nonumber\\
&& + 
  (  {\frac{72\,\epsilon \,{{\lambda }^4}}{{{\tau }_{-1 + m}}}} + 
     {\frac{72\,\epsilon \,{{\lambda }^4}}{{{\tau }_m}}} 
  )\,C^{(6)}_{-1 + m,m,1 + m}   +
     {\frac{72\,\epsilon \,{{\lambda }^4}}{{{\tau }_m}}}
   \,C^{(6)}_{1 + m,1 + m,m}+\nonumber\\
&&  + 
     {\frac{72\,\epsilon \,{{\lambda }^4}}{{{\tau }_{-1 + m}}}}
   \,C^{(6)}_{-1 + m,-1 + m,m}  + 
     {\frac{9\,\epsilon \,{{\lambda }^4}}{{{\tau }_{1 + m}}}}
   \,C^{(6)}_{m,m,2 + m} +
     {\frac{9\,\epsilon \,{{\lambda }^4}}{{{\tau }_{-2 + m}}}}
   \,C^{(6)}_{m,m,-2 + m}
\nonumber
\end{eqnarray}
\begin{eqnarray}
&&0=\sum_{p,q,r} [I_{m,m,m-1;p,q,r}^{(6;0)}+
\epsilon I_{m,m,m-1;p,q,r}^{(6;1)}]\,C_{p,q,r}^{(6)}= \nonumber\\
&&=   
  (  {\frac{{{\lambda }^2}}{{{\tau }_{-1 + m}}}} - 
     {\frac{5\,\epsilon \,{{\lambda }^4}}{{{\tau }_m}}} - 
     {\frac{23\,\epsilon \,{{\lambda }^4}}{{{\tau }_{-1 + m}}}} 
  )\,C^{(6)}_{m,m,m}    + 
  (  {\frac{10\,\epsilon \,{{\lambda }^4}}{{{\tau }_{-1 + m}}}} + 
     {\frac{15\,\epsilon \,{{\lambda }^4}}{{{\tau }_m}}} 
  )\,C^{(6)}_{m,m,1 + m}     +\nonumber\\
&&       - 
  (  {\frac{{{\lambda }^2}}{{{\tau }_{-2 + m}}}} + 
     {\frac{2\,{{\lambda }^2}}{{{\tau }_m}}} + 
     {\frac{7\,{{\lambda }^2}}{{{\tau }_{-1 + m}}}} - 
     {\frac{17\,\epsilon \,{{\lambda }^4}}{{{\tau }_{-2 + m}}}} - 
     {\frac{40\,\epsilon \,{{\lambda }^4}}{{{\tau }_m}}} - 
     {\frac{169\,\epsilon \,{{\lambda }^4}}{{{\tau }_{-1 + m}}}} 
  )\,C^{(6)}_{m,m,-1 + m}  +\nonumber\\
&&   + 
  (  {\frac{4\,{{\lambda }^2}}{{{\tau }_m}}}    - 
     {\frac{4\,\epsilon \,{{\lambda }^4}}{{{\tau }_{1 + m}}}}    -  
     {\frac{8\,\epsilon \,{{\lambda }^4}}{{{\tau }_{-2 + m}}}}   - 
     {\frac{36\,\epsilon \,{{\lambda }^4}}{{{\tau }_{-1 + m}}}}  - 
     {\frac{96\,\epsilon \,{{\lambda }^4}}{{{\tau }_m}}}
  )\,C^{(6)}_{-1 + m,m,1 + m}  +\nonumber\\
&&   + 
  (  {\frac{4\,{{\lambda }^2}}{{{\tau }_{-1 + m}}}}  - 
     {\frac{8\,\epsilon \,{{\lambda }^4}}{{{\tau }_m}}}  - 
     {\frac{12\,\epsilon \,{{\lambda }^4}}{{{\tau }_{-2 + m}}}}  - 
     {\frac{124\,\epsilon \,{{\lambda }^4}}{{{\tau }_{-1 + m}}}} 
  )\,C^{(6)}_{-1 + m,-1 + m,m} +\nonumber\\
&&    + 
  (  {\frac{8\,\epsilon \,{{\lambda }^4}}{{{\tau }_{-1 + m}}}} + 
     {\frac{8\,\epsilon \,{{\lambda }^4}}{{{\tau }_m}}} 
  )\,C^{(6)}_{-1 + m,-1 + m,1 + m}     + 
     {\frac{8\,\epsilon \,{{\lambda }^4}\,}{{{\tau }_m}}}
     C^{(6)}_{1 + m,1 + m,-1 + m} +\nonumber\\
&&   + 
     {\frac{8\,\epsilon \,{{\lambda }^4}}{{{\tau }_{-1 + m}}}} 
   \,C^{(6)}_{-1 + m,-1 + m,-1 + m}   + 
  (  {\frac{12\,\epsilon \,{{\lambda }^4}}{{{\tau }_{-1 + m}}}}   +
     {\frac{24\,\epsilon \,{{\lambda }^4}}{{{\tau }_{-2 + m}}}}
  )\,C^{(6)}_{-2 + m,-1 + m,m} +\nonumber\\
&&  +
 (   {\frac{{{\lambda }^2}}{{{\tau }_{-2 + m}}}}- 
     {\frac{\epsilon \,{{\lambda }^4}}{{{\tau }_{-3 + m}}}} - 
     {\frac{4\,\epsilon \,{{\lambda }^4}}{{{\tau }_m}}}  - 
     {\frac{6\,\epsilon \,{{\lambda }^4}}{{{\tau }_{-1 + m}}}} - 
     {\frac{17\,\epsilon \,{{\lambda }^4}}{{{\tau }_{-2 + m}}}}
  )\,C^{(6)}_{m,m,-2 + m}     +\nonumber\\
&&  +   
  (  {\frac{8\,\epsilon \,{{\lambda }^4}}{{{\tau }_{-2 + m}}}} + 
     {\frac{8\,\epsilon \,{{\lambda }^4}}{{{\tau }_m}}} 
  )\,C^{(6)}_{-2 + m,m,1 + m}        + 
     {\frac{4\,\epsilon \lambda^4}{\tau_{1 + m}}}
   \,C^{(6)}_{-1 + m,m,2 + m}  +
     {\frac{\epsilon \,{{\lambda }^4}}{{{\tau }_{-3 + m}}}} 
   \,C^{(6)}_{m,m,-3 + m} 
\nonumber 
\end{eqnarray}
\begin{eqnarray}
&&\sum_{p,q,r}[I_{m,m-1,m-1;p,q,r}^{(6;0)}+\epsilon 
I_{m,m-1,m-1;p,q,r}^{(6;1)}]\,C_{p,q,r}^{(6)}= \nonumber\\
&&=   
     {\frac{8\,\epsilon \,{{\lambda }^4}}{{{\tau }_{-1 + m}}}}
   \,C^{(6)}_{m,m,m}  + 
  (  {\frac{4\,{{\lambda }^2}}{{{\tau }_{-1 + m}}}}  -
     {\frac{8\,\epsilon \,{{\lambda }^4}}{{{\tau }_{-2 + m}}}} - 
     {\frac{12\,\epsilon \,{{\lambda }^4}}{{{\tau }_m}}} - 
     {\frac{124\,\epsilon \,{{\lambda }^4}}{{{\tau }_{-1 + m}}}} 
  )\,C^{(6)}_{m,m,-1 + m}  +\nonumber\\
&& +     
  (  {\frac{12\,\epsilon \,{{\lambda }^4}}{{{\tau }_{-1 + m}}}} + 
     {\frac{24\,\epsilon \,{{\lambda }^4}}{{{\tau }_m}}} 
  )\,C^{(6)}_{-1 + m,m,1 + m} +\nonumber\\
&& -   
  (  {\frac{{{\lambda }^2}}{{{\tau }_m}}} +
     {\frac{2\,{{\lambda }^2}}{{{\tau }_{-2 + m}}}} + 
     {\frac{7\,{{\lambda }^2}}{{{\tau }_{-1 + m}}}} - 
     {\frac{17\,\epsilon \,{{\lambda }^4}}{{{\tau }_m}}} - 
     {\frac{40\,\epsilon \,{{\lambda }^4}}{{{\tau }_{-2 + m}}}}   - 
     {\frac{169\,\epsilon \,{{\lambda }^4}}{{{\tau }_{-1 + m}}}}
  )\,C^{(6)}_{-1 + m,-1 + m,m}    +\nonumber\\
&&  +  
  (  {\frac{{{\lambda }^2}}{{{\tau }_m}}}  -
     {\frac{4\,\epsilon \,{{\lambda }^4}}{{{\tau }_{-2 + m}}}} - 
     {\frac{6\,\epsilon \,{{\lambda }^4}}{{{\tau }_{-1 + m}}}} - 
     {\frac{17\,\epsilon \,{{\lambda }^4}}{{{\tau }_m}}} - 
     {\frac{\epsilon \,{{\lambda }^4}}{{{\tau }_{1 + m}}}}
  )\,C^{(6)}_{-1 + m,-1 + m,1 + m}  +\nonumber\\
&&  + 
 (  {\frac{{{\lambda }^2}}{{{\tau }_{-1 + m}}}}  -
    {\frac{5\,\epsilon \,{{\lambda }^4}}{{{\tau }_{-2 + m}}}} - 
    {\frac{23\,\epsilon \,{{\lambda }^4}}{{{\tau }_{-1 + m}}}}
 )\,C^{(6)}_{-1 + m,-1 + m,-1 + m}  +\nonumber\\
&& + 
 (  {\frac{8\,\epsilon \,{{\lambda }^4}}{{{\tau }_{-2 + m}}}} + 
    {\frac{8\,\epsilon \,{{\lambda }^4}}{{{\tau }_{-1 + m}}}}
 )\,C^{(6)}_{m,m,-2 + m}     + 
 (   {\frac{15\,\epsilon \,{{\lambda }^4}}{{{\tau }_{-2 + m}}}} + 
     {\frac{10\,\epsilon \,{{\lambda }^4}}{{{\tau }_{-1 + m}}}}
 )\,C^{(6)}_{-1 + m,-1 + m,-2 + m}  +\nonumber\\
&& +   
  (  {\frac{4\,{{\lambda }^2}}{{{\tau }_{-2 + m}}}}   - 
     {\frac{4\,\epsilon \,{{\lambda }^4}}{{{\tau }_{-3 + m}}}} - 
     {\frac{8\,\epsilon \,{{\lambda }^4}}{{{\tau }_m}}}  - 
     {\frac{36\,\epsilon \,{{\lambda }^4}}{{{\tau }_{-1 + m}}}} - 
     {\frac{96\,\epsilon \,{{\lambda }^4}}{{{\tau }_{-2 + m}}}}
  )\,C^{(6)}_{-2 + m,-1 + m,m} +\nonumber\\
&&  + 
     {\frac{\epsilon \lambda^4}{\tau_{1 + m}}}
   \,C^{(6)}_{-1 + m,-1 + m,2 + m}  +   
  (  {\frac{8\,\epsilon \,{{\lambda }^4}}{{{\tau }_{-2 + m}}}} + 
     {\frac{8\,\epsilon \,{{\lambda }^4}}{{{\tau }_m}}} 
  )\,C^{(6)}_{-2 + m,-1 + m,1 + m} +\nonumber\\
&& + 
    {\frac{8\,\epsilon \,{{\lambda }^4}}{{{\tau }_{-2 + m}}}}
  \,C^{(6)}_{-2 + m,-2 + m,m} +
    {\frac{4\,\epsilon \,{{\lambda }^4}}{{{\tau }_{-3 + m}}}}
  \,C^{(6)}_{-3 + m,-1 + m,m}
\nonumber
\end{eqnarray}
\begin{eqnarray}
&&\sum_{p.q,r}[I_{m,m+1,m-1;p,q,r}^{(6;0)}+ \epsilon
I_{m,m+1,m-1;p,q,r}^{(6;1)}]C_{p,q,r}^{(6)}=\nonumber\\
&&=  
 (  {\frac{2\,\epsilon \,{{\lambda }^4}}{{{\tau }_{-1 + m}}}} + 
    {\frac{2\,\epsilon \,{{\lambda }^4}}{{{\tau }_m}}} 
 )\,C^{(6)}_{m,m,m}        + 
 (  {\frac{{{\lambda }^2}}{{{\tau }_{-1 + m}}}} - 
    {\frac{2\,\epsilon \,{{\lambda }^4}}{{{\tau }_{1 + m}}}} - 
    {\frac{12\,\epsilon \,{{\lambda }^4}}{{{\tau }_m}}} - 
    {\frac{22\,\epsilon \,{{\lambda }^4}}{{{\tau }_{-1 + m}}}}
 )\,C^{(6)}_{m,m,1 + m}   +\nonumber\\
&&  + 
 (  {\frac{{{\lambda }^2}}{{{\tau }_m}}}    - 
    {\frac{2\,\epsilon \,{{\lambda }^4}}{{{\tau }_{-2 + m}}}} - 
    {\frac{12\,\epsilon \,{{\lambda }^4}}{{{\tau }_{-1 + m}}}} - 
    {\frac{22\,\epsilon \,{{\lambda }^4}}{{{\tau }_m}}} 
 )\,C^{(6)}_{m,m,-1 + m}     +\nonumber\\
&&  + 
 (  {\frac{3\,\epsilon \,{{\lambda }^4}}{{{\tau }_{-1 + m}}}} + 
    {\frac{6\,\epsilon \,{{\lambda }^4}}{{{\tau }_m}}} 
 )\,C^{(6)}_{1 + m,1 + m,m}   + 
 (  {\frac{3\,\epsilon \,{{\lambda }^4}}{{{\tau }_m}}}   +
    {\frac{6\,\epsilon \,{{\lambda }^4}}{{{\tau }_{-1 + m}}}}
 )\,C^{(6)}_{-1 + m,-1 + m,m}    +\nonumber\\
&& - 
 (  {\frac{{{\lambda }^2}}{{{\tau }_{-2 + m}}}}  + 
    {\frac{{{\lambda }^2}}{{{\tau }_{1 + m}}}} + 
    {\frac{4\,{{\lambda }^2}}{{{\tau }_{-1 + m}}}}  + 
    {\frac{4\,{{\lambda }^2}}{{{\tau }_m}}}    - 
    {\frac{18\,\epsilon \,{{\lambda }^4}}{{{\tau }_{-2 + m}}}}- 
    {\frac{18\,\epsilon \,{{\lambda }^4}}{{{\tau }_{1 + m}}}} - 
    {\frac{84\,\epsilon \,{{\lambda }^4}}{{{\tau }_{-1 + m}}}} - 
    {\frac{84\,\epsilon \,{{\lambda }^4}}{{{\tau }_m}}} 
 )\,C^{(6)}_{-1 + m,m,1 + m}      +\nonumber\\
&&  + 
 (  {\frac{{{\lambda }^2}}{{{\tau }_m}}}   -
    {\frac{2\,\epsilon \,{{\lambda }^4}}{{{\tau }_{-2 + m}}}} - 
    {\frac{3\,\epsilon \,{{\lambda }^4}}{{{\tau }_{-1 + m}}}} - 
    {\frac{3\,\epsilon \,{{\lambda }^4}}{{{\tau }_{1 + m}}}}  - 
    {\frac{20\,\epsilon \,{{\lambda }^4}}{{{\tau }_m}}}
 )\,C^{(6)}_{1 + m,1 + m,-1 + m}  +\nonumber\\
&& + 
 (  {\frac{{{\lambda }^2}}{{{\tau }_{-1 + m}}}}   - 
    {\frac{2\,\epsilon \,{{\lambda }^4}}{{{\tau }_{1 + m}}}} -
    {\frac{3\,\epsilon \,{{\lambda }^4}}{{{\tau }_{-2 + m}}}} - 
    {\frac{3\,\epsilon \,{{\lambda }^4}}{{{\tau }_m}}} - 
    {\frac{20\,\epsilon \,{{\lambda }^4}}{{{\tau }_{-1 + m}}}}
 )\,C^{(6)}_{-1 + m,-1 + m,1 + m}   +\nonumber\\
&&  + 
 (  {\frac{3\,\epsilon \,{{\lambda }^4}}{{{\tau }_m}}} + 
    {\frac{6\,\epsilon \,{{\lambda }^4}}{{{\tau }_{1 + m}}}}
 )\,C^{(6)}_{-1 + m,1 + m,2 + m}  + 
 (  {\frac{3\,\epsilon \,{{\lambda }^4}}{{{\tau }_{-1 + m}}}} +
    {\frac{6\,\epsilon \,{{\lambda }^4}}{{{\tau }_{-2 + m}}}}
 )\,C^{(6)}_{-2 + m,-1 + m,1 + m}    +\nonumber\\
&& + 
 (  {\frac{2\,\epsilon \,{{\lambda }^4}}{{{\tau }_{-1 + m}}}} + 
    {\frac{2\,\epsilon \,{{\lambda }^4}}{{{\tau }_{1 + m}}}}
 )\,C^{(6)}_{-1 + m,-1 + m,2 + m} + 
 (  {\frac{2\,\epsilon \,{{\lambda }^4}}{{{\tau }_{-1 + m}}}} + 
    {\frac{2\,\epsilon \,{{\lambda }^4}}{{{\tau }_{1 + m}}}}
 )\,C^{(6)}_{m,m,2 + m} +\nonumber\\
&& + 
 (  {\frac{2\,\epsilon \,{{\lambda }^4}}{{{\tau }_{-2 + m}}}} + 
    {\frac{2\,\epsilon \,{{\lambda }^4}}{{{\tau }_m}}} 
 )\,C^{(6)}_{m,m,-2 + m}    + 
 (  {\frac{2\,\epsilon \,{{\lambda }^4}}{{{\tau }_{-2 + m}}}} + 
    {\frac{2\,\epsilon \,{{\lambda }^4}}{{{\tau }_m}}} 
 )\,C^{(6)}_{1 + m,1 + m,-2 + m}  +\nonumber\\
&& + 
 (  {\frac{{{\lambda }^2}}{{{\tau }_{-2 + m}}}}    -  
    {\frac{\epsilon \,{{\lambda }^4}}{{{\tau }_{-3 + m}}}} - 
    {\frac{2\,\epsilon \,{{\lambda }^4}}{{{\tau }_{1 + m}}}} - 
    {\frac{3\,\epsilon \,{{\lambda }^4}}{{{\tau }_{-1 + m}}}} - 
    {\frac{8\,\epsilon \,{{\lambda }^4}}{{{\tau }_m}}} - 
    {\frac{18\,\epsilon \,{{\lambda }^4}}{{{\tau }_{-2 + m}}}}
 )\,C^{(6)}_{-2 + m,m,1 + m}      +\nonumber\\
&&  + 
 (  {\frac{{{\lambda }^2}}{{{\tau }_{1 + m}}}}  - 
    {\frac{\epsilon \,{{\lambda }^4}}{{{\tau }_{2 + m}}}}  -
    {\frac{2\,\epsilon \,{{\lambda }^4}}{{{\tau }_{-2 + m}}}} - 
    {\frac{3\,\epsilon \,{{\lambda }^4}}{{{\tau }_m}}} - 
    {\frac{8\,\epsilon \,{{\lambda }^4}}{{{\tau }_{-1 + m}}}} - 
    {\frac{18\,\epsilon \,{{\lambda }^4}}{{{\tau }_{1 + m}}}}
 )\,C^{(6)}_{-1 + m,m,2 + m} +\nonumber\\
&& + 
 (  {\frac{2\,\epsilon \,{{\lambda }^4}}{{{\tau }_{-2 + m}}}} + 
    {\frac{2\,\epsilon \,{{\lambda }^4}}{{{\tau }_{1 + m}}}}
 )\,C^{(6)}_{-2 + m,m,2 + m}    + 
    {\frac{\epsilon \,{{\lambda }^4}}{{{\tau }_{2 + m}}}}
  \,C^{(6)}_{-1 + m,m,3 + m} +
    {\frac{\epsilon \,{{\lambda }^4}}{{{\tau }_{-3 + m}}}}
  \,C^{(6)}_{-3 + m,m,1 + m}
\label{6closedexplicit}
\end{eqnarray}

\section{The inertial operator for the eight moment of the correlation up 
         to first order} 
\label{ap:eightorder}

The set of independent equations is finally given as: 
\begin{equation}
  \left\{%
    \begin{array}{l}
\sum_{p,q,r,s}[I_{m,m,m,m;p,q,r,s}^{(8;0)}+\epsilon 
I_{m,m,m,m;p,q,r,s}^{(8;1)}] C_{p,q,r,s}^{(8)}=0\nonumber\\
\sum_{p,q,r,s}[I_{m,m,m,m-1;p,q,r,s}^{(8,0)}+\epsilon 
I_{m,m,m,m-1;p,q,r,s}^{(8;1)}] C_{p,q,r,s}^{(8)}=0\nonumber\\
\sum_{p,q,r,s}[I_{m,m,m-1,m-1;p,q,r,s}^{(8,0)} + \epsilon 
I_{m,m,m-1,m-1;p,q,r,s}^{(8;1)}]C_{p,q,r,s}^{(8)}=0 \nonumber\\
\sum_{p,q,r,s}[I_{m,m,m+1,m-1;p,q,r,s}^{(8,0)} +\epsilon 
I_{m,m,m+1,m-1;p,q,r,s}^{(8;1)}] C_{p,q,r,s}^{(8)}=0\nonumber\\
\sum_{p,q,r,s}[I_{m,m-1,m-1,m-1;p,q,r,s}^{(8,0)}+\epsilon 
I_{m,m-1,m-1,m-1;p,q,r,s}^{(8;1)}] C_{p,q,r,s}^{(8)}=0 \nonumber\\
\sum_{p,q,r,s}[I_{m,m+1,m-1,m-1;p,q,r,s}^{(8,0)}+\epsilon 
I_{m,m+1,m-1,m-1;p,q,r,s}^{(8;1)}]C_{p,q,r,s}^{(8)}=0 \nonumber\\
\sum_{p,q,r,s}[I_{m,m+1,m+1,m-1;p,q,r,s}^{(8,0)}+\epsilon 
I_{m,m+1,m+1,m-1;p,q,r,s}^{(8;1)}]C_{p,q,r,s}^{(8)}=0 \nonumber\\
\sum_{p,q,r,s}[I_{m,m+1,m-1,m-2;p,q,r,s}^{(8,0)}+\epsilon 
I_{m,m+1,m-1,m-2;p,q,r,s}^{(8;1)}] C_{p,q,r,s}^{(8)}=0 
\label{8closed}
    \end{array}
  \right.
\end{equation}
The closure is provided again assuming scaling for all the possible
conditioned expectation values with respect to a given shell. It follows
\begin{eqnarray}
C_{m+n,m+n,m+n,m+n}^{(8)}&=&z(\epsilon)^{-l}C_{m,m,m,m}^{(8)}\nonumber\\
C_{m+n,m+n,m+n,m}^{(8)}&=&y_{1}(\epsilon) k^{-H(6,\epsilon)}_{n-1}  
C_{m,m,m,m}^{(8)}\nonumber\\
C_{m+n,m+n,m,m}^{(8)}&=&y_{2 }(\epsilon)k^{- H(4,\epsilon)}_{n-1} 
C_{m,m,m,m}^{(8)}\nonumber\\
C_{m+n,m,m,m}^{(8)}&=&y_{3}(\epsilon)k^{- H(2)}_{n-1} 
C_{m,m,m,m}^{(8)} \nonumber\\
C_{m+n+p,m+n,m,m}^{(8)}&=&y_{4}(\epsilon) k^{-H(2)}_{p-1} 
k^{- H(4,\epsilon)}_{n-1}  C_{m,m,m,m}^{(8)}\nonumber\\
C_{m+n+p,m+n,m+n,m}^{(8)}&=& y_{5}(\epsilon) k^{-H(2)}_{p-1} 
k^{- H(6,\epsilon)}_{n-1} C_{m,m,m,m}^{(8)}\nonumber\\
C_{m+n+p,m+n+p,m+n,m}^{(8)}&=&y_{6}(\epsilon) k^{- H(4,\epsilon)}_{p-1}
k^{- H(6,\epsilon)}_{n-1} C_{m,m,m,m}^{(8)}\nonumber\\
C_{m+n+p+q,m+n+p,m+n,m}^{(8)}&=&y_{7}(\epsilon) k^{- H(2)}_{q-1}
k^{- H(4,\epsilon)}_{p-1} k^{- H(6,\epsilon)}_{n-1} C_{m,m,m,m}^{(8)}
\label{8scaling}
\end{eqnarray}

\newpage


\begin{thebibliography}{99}
\bibitem{Frisch}
U.Frisch, {\it Turbulence: the legacy of A.N. Kolmogorov} 
(Cambridge University Press 1995)
\bibitem{Kraichnan}
R.H. Kraichnan, Phys. Rev. Lett. {\bf 72}, 1016 (1994)
\bibitem{BeGaKu}
D.Bernard, K.Gawedzki and A.Kupiainen Phys. Rev. {\bf E 54} 2564 (1996)
\bibitem{GaKu}
K. Gawedzki and A. Kupiainen, Phys. Rev. Lett. {\bf 75} 3834 (1995)
\bibitem{CFKL}
M. Chertkov, G. Falkovich, I. Kolokov and V. Lebedev, Phys. Rev. {\bf E 52} 4924
(1995)
\bibitem{CF}
M.Chertkov and G.Falkovich, Phys. Rev. Lett. {\bf 76} 2706 (1996)
\bibitem{Ga}
K.Gawedzki, chao-dyn/9803027
\bibitem{FMV}
U.Frisch, A.Mazzino and M.Vergassola, Phys. Rev. Lett. {\bf 80}, 5532 (1998)
and cond-mat/9802192
\bibitem{Vulpio}
M.H. Jensen, G. Paladin and A. Vulpiani, Phys. Rev.{\bf A 45}, N. 10, 
7214 (1992)  
\bibitem{libro}
T.Bohr, M.H. Jensen, G.Paladin and A.Vulpiani 
{\it Dynamical Systems Approach to Turbulence} 
Cambridge University Press (1998)
\bibitem{Obukhov} 
A.M. Obukhov, Ivz. Akad. SSSR, Serv Geogr. Geofiz. {\bf 13}, 58 (1949)
\bibitem{Corrsin}
S. Corrsin, J.Appl. Phys. {\bf 22}, 469 (1951)
\bibitem{Erdelyi}
A.Erd\`elyi {\it Asymptotic expansions} 
(Dover Publicatins Inc. New York 1956)
\bibitem{Chertkov}
M.Chertkov, G.Falkovich and V.Lebedev Phys.Rev.Lett. {\bf 76}, 3707 (1996)
\bibitem{Arnold}
L.Arnold: 
{\it Stochastic differential equations: Theory and 
Applications} (Wiley, New York 1974)
\bibitem{WB}
A.Wirth and L.Biferale, Phys.Rev. E {\bf 54}, 4982 (1996)
\bibitem{BBW}
R. Benzi, L.Biferale and A.Wirth, Phys. Rev. Lett. {\bf 78}, p. 4926 (1997)  . 
\bibitem{Bass}
R.F.Bass, {\it Diffusions and elliptic operators} 
New York Springer-Verlag 1998  
\bibitem{Nualart}
D.Nualart, {\it The Malliavin calculus and related topics} 
New York London Springer-Verlag 1995 
\bibitem{Cardy}
J.Cardy {\it Scaling and Renormalization in Statistical Physics}
Cambridge Lecture Notes in Physics, Cambridge University Press (1996)
\bibitem{Zinn}
J Zinn-Justin {\it Quantum Field Theory and Critical Phenomena}
Clarendon Press Oxford (1989)
\bibitem{Richardson}
L.F. Richardson {\it Weather prediction by Numerical Process} Cambridge 
University Press, Cambridge (1922)
\bibitem{Parisi}
G. Parisi and U.Frisch {\it Turbulence and Predictability in Geophysical Fluid 
Dynamics}, Proceed. Intern. School of Physics ``E.Fermi'' Varenna 1983 
eds M.Ghil, R.Benzi and G.Parisi. North Holland, Amsterdam.
\bibitem{multi}
L.Biferale, G.Boffetta, A. Celani and F.Toschi chao-dyn/9804035
\bibitem{LeBellac}
M.Le Bellac {\it Quantum and statistical field theory}
Oxford University Press 1991
\bibitem{Kloeden}
P.E.Kloeden and E.Platen {\it Numerical Solution of Stochastic Differential 
Equations} Springer-Verlag 1995 
\bibitem{Burrage}
k.Burrage and P.Burrage preprint
{\it High strong order methods for non commutative ordinary stochastic 
differential equations and the Magnus formula} 
http://www.maths.uq.edu.au/~kb/papers.html
\bibitem{Sidje}
R.B.Sidje, to appear in ACM-Transactions on Mathematical softwere
\bibitem{SanMiguel}
M.San Miguel and R.Toral in 
{\it Instabilities and Non\-equilibrium Structures VI} 
E.Tirapegui and W.Zeller editors Kluwer Acad. Pub (1997) and
cond-mat/9707147
\bibitem{ESS}
R.Benzi, S.Ciliberto, R.Tripiccione, C Baudet and S.Succi Phys.Rev E. {\bf 48}
R29 (1993)
\end{thebibliography}
\end{document}